\documentclass[universe,review,accept,pdftex,moreauthors]{Definitions/mdpi} 

\firstpage{1} 
\pubvolume{10}
\issuenum{1}
\articlenumber{0}
\pubyear{2024}
\copyrightyear{2024}
\externaleditor{Academic Editor: Firstname \linebreak  Lastname}
\datereceived{15 January 2024} 
\daterevised{6 February 2024} 
\dateaccepted{7 February 2024} 
\datepublished{ } 
\hreflink{https://doi.org/} 

\Title{Recent Findings from {Heavy-Flavor} 
	 Angular Correlation Measurements in Hadronic Collisions}

\TitleCitation{Recent Findings from Heavy-Flavor Angular Correlation Measurements in Hadronic Collisions}

\Author{Deepa Thomas 
 $^{1,}$*
\orcidA{} and Fabio Colamaria $^{2,}$*\orcidB{}}

\AuthorNames{Deepa Thomas and Fabio Colamaria}

\AuthorCitation{Thomas, D.
; Colamaria, F.}

\address{%
$^{1}$ \quad The University 
 of Texas at Austin, Department of Physics, 2515 Speedway, Austin, TX 78712, USA 
\\
$^{2}$ \quad Istituto Nazionale di Fisica Nucleare, 
 Sezione di Bari, Via E. Orabona, 4 - 70125 Bari, Italy}

\corres{{Correspondence: 
 deepa.thomas@utexas.edu (D.T.); fabio.colamaria@ba.infn.it (F.C.)}}

\abstract{The study of angular correlations of heavy-flavor particles in hadronic collisions can provide crucial insight into the heavy quark production, showering, and hadronization processes. The comparison with model predictions allows us to discriminate among different approaches for heavy quark production and hadronization, as well as different treatments of the underlying event    employed by the models to reproduce correlation observables.
In ultra-relativistic heavy-ion collisions, where a deconfined state of matter, the quark--gluon plasma (QGP), is created, heavy-flavor correlations can shed light on the modification of the heavy quark fragmentation due to the interaction between charm and beauty quarks with the QGP constituents, as well as characterize their energy loss processes while traversing the medium. Insight into the possible emergence of collective-like mechanisms in smaller systems, resembling those observed in heavy-ion collisions, can also be obtained by performing correlation studies in high-multiplicity proton--proton and proton--nucleus collisions.
In this review, the most recent and relevant measurements of heavy-flavor correlations performed in all collision systems at the LHC and RHIC will be presented, and the new understandings that they provide will be discussed.}


\newcommand{\pt}{p_{\rm T}}
\newcommand{\dphi}{\Delta\varphi}
\newcommand{\deta}{\Delta\eta}
\newcommand{\Dzero}{{\rm D^0}}
\newcommand{\Dstar}{{\rm D^{*+}}}
\newcommand{\Dplus}{{\rm D^+}}

\newcommand{\s}{\sqrt{s}}
\newcommand{\sNN}{\sqrt{s_{\rm NN}}}

\usepackage{comment}
\begin{document}


\section{Introduction}

The study of heavy quark (charm and beauty quarks) production in high-energy hadronic collisions is an important tool to test and validate perturbative quantum chromodynamics (pQCD) calculations~\cite{Cacciari:1998it, Cacciari:2001td, Kniehl:2005ej, Kniehl:2007erq}, as they are produced in hard parton scattering processes. The production cross-section of several heavy-flavor hadrons and their decay products has been measured at different centers of mass energies at RHIC~\cite{Aggarwal:2010xp, Adare:2009ic,PHENIX:2006tli}, \mbox{Tevatron~\cite{CDF:2003vmf,Cacciari:2003zu,Acosta:2004yw},} and at the LHC~\cite{ALICE:2019nuy,CMS:2017qjw,LHCb:2015swx,ALICE:2021wct,ALICE:2022mur,ALICE:2019nxm,ALICE:2019rmo,ALICE:2021mgk,ALICE:2012msv,ALICE:2012inj,ALICE:2012mzy,Aaij:2016jht,ALICE:2012acz,ALICE:2014ivb, Aaij:2012jd, Aad:2012jga, ATLAS:2013cia, Chatrchyan:2012hw, Khachatryan:2011mk, Chatrchyan:2011pw, Chatrchyan:2011vh, Khachatryan:2016csy}, and are compared with pQCD calculations~\cite{Cacciari:2012ny, Kniehl:2008eu,Kniehl:2011bk,Kniehl:2005ej,Bolzoni:2012kx,Bolzoni_2014}. The correlated production of heavy flavors, studied as a function of variables such as the azimuthal angle between heavy-flavor particles, either by direct reconstruction of heavy-flavor hadrons or from their decay products, can provide a significantly larger amount of information than single-particle inclusive heavy-flavor production.  

In ultra-relativistic heavy-ion collisions, heavy quarks play an important role in the study of the deconfined phase of strongly interacting matter, the quark--gluon plasma (QGP), created in these collisions. Traditional observables of heavy quarks, such as the nuclear modification factor ($R_{\rm{AA}}$) and the elliptic flow coefficient ($v_2$), have been extensively studied at RHIC~\cite{PHENIX:2010xji, STAR:2006btx, STAR:2017kkh} and at the LHC~\cite{ALICE:2021rxa, CMS:2017qjw, ALICE:2021kfc, ALICE:2017pbx, ALICE:2020iug}. These measurements indicate that heavy quarks experience significant in-medium energy loss at large transverse momentum ($p_{\rm{T}}$), and that charm quarks partially thermalize within the medium at smaller $p_{\rm{T}}$. These effects are induced by the interaction of heavy quarks with the medium constituents, mainly made up of light partons. The interaction has two main contributions, the purely elastic process, resulting in a collisional energy loss~\cite{Braaten:1991jj,Peshier:2006hi,Peigne:2008nd}, and gluon bremsstrahlung, producing a radiative energy loss~\cite{Gyulassy:1993hr,Baier:1994bd,Gyulassy:2000fs,Dokshitzer:2001zm,Armesto:2004vz,Zhang:2003wk}. Both of these processes depend on the parton mass; thus, studies of heavy quark in heavy-ion collisions provide important information about the properties of the QGP. Currently, it remains a challenge to describe $R_{\rm{AA}}$ and $v_2$ simultaneously from a low to high range of transverse momentum in a given theoretical framework. More differential observables, such as angular correlations between heavy-flavor particles, are more sensitive to the specific interaction processes between heavy quarks and the QGP constituents, and can thus provide further information on the propagation of heavy quarks in the QGP medium~\cite{Nahrgang:2013saa, Cao:2015cba}.

In proton--nucleus (p--A) collisions, due to the presence of the nucleus in the initial state, several cold-nuclear-matter effects can influence the production, fragmentation, and hadronization of heavy quarks, such as the impact of the nuclear parton distribution function (nPDF)~\cite{Eskola:2009uj,deFlorian:2003qf,Hirai:2007sx}, the presence of a coherent and saturated gluonic system dominated by gluons at low Bjorken-$x$~\cite{Fujii:2013yja,Tribedy:2011aa,Albacete:2012xq,Rezaeian:2012ye}, and partons undergoing multiple elastic, inelastic, and coherent scatterings~\cite{Accardi:2009qv,Salgado:2011wc}. These effects can also modify the heavy-flavor correlation distribution~\cite{Vogt:2018oje, Vogt:2019xmm, Marquet:2017xwy}, and their understanding is crucial for the interpretation of any modification of the correlation distributions in the presence of a QGP, produced in heavy-ion~collisions. 

In this article, we review and summarize the latest measurements of angular correlations of heavy-flavor particles in proton--proton (pp), proton--nucleus (p--A), and nucleus--nucleus (A--A) collisions at the LHC and RHIC energies. The article is structured as follows. In Section~\ref{sec:ProdMechPp}, studies of heavy-flavor production mechanisms in pp and p--A collisions using angular correlation techniques are discussed. The fragmentation of heavy quarks into final-state jets in pp and p--A collisions was studied using angular correlations of heavy-flavor hadrons and charged particles, as discussed in Section~\ref{sect:jetfragm}. In Section~\ref{sect:corrHIC}, new insight into heavy quark propagation and energy loss in the QGP obtained using correlations of heavy quarks is discussed. In recent years, questions have been raised about the possible formation of a QGP in smaller collision systems, such as in pp and p--A collisions, due to the observation of a ``long-range ridge structure'' in two-particle azimuthal correlations of light-flavor particles. To further investigate this possibility, measurements of azimuthal correlations of heavy-flavor particles and charged hadrons were performed, from which the azimuthal anisotropy of heavy-flavor particles was extracted. These studies are reviewed in Section~\ref{sect:collectivity}. 

\section{Study of Heavy Quark Production Mechanisms}
\label{sec:ProdMechPp}
In high-energy hadronic collisions, heavy quarks (charm and beauty) are mainly produced in hard parton scattering processes. An inclusive production of several heavy-flavor hadrons has been measured in experiments at the LHC and at lower energies, and is compared with pQCD calculations such as FONLL and GM-VFNS~\cite{Cacciari:2012ny, Kniehl:2008eu,Kniehl:2011bk,Kniehl:2005ej,Bolzoni:2012kx,Bolzoni_2014}. Exclusive measurements of heavy-flavor correlations, e.g., as a function of the azimuthal angle between heavy-flavor hadrons, $\varphi$, are a stronger test of the heavy quark pair (${\rm Q\bar{Q}}$) production mechanisms than single-particle inclusive distributions. At leading order (LO), $2 \rightarrow 2$ in $\alpha_{\rm s}$ for the parton interaction sub-processes, the heavy quark pairs will be emitted with a back-to-back topology in azimuth due to momentum conservation. At next-to-leading order (NLO), $2 \rightarrow 2+n~(n \geq 1)$, additional partons are emitted, resulting in different topologies of the produced heavy quarks~\cite{Vogt:2018oje, Vogt:2019xmm}. Experimentally, correlation patterns of heavy quarks can be {assessed} 
with measurements of angular correlations of heavy-flavor particle pairs, as discussed below. 

Measurements of angular correlations of heavy-flavor particles in proton--proton (pp) collisions that fit this context were performed by the PHENIX Collaboration at RHIC at $\sqrt{s}= 200$ GeV~\cite{PHENIX:2018dwt}, and by ATLAS~\cite{ATLAS:2017wfq}, CMS~\cite{CMS:2011yuk}, and LHCb~\cite{LHCb:2012aiv, LHCb:2017bvf} Collaborations at the LHC at $\sqrt{s}= 7$ and 8 TeV, and in proton--lead (p--Pb) collisions at $\sqrt{s_{\rm{NN}}}=8.16$ TeV by the LHCb Collaboration~\cite{LHCb:2020jse}. The goal of these measurements is to understand the correlated production of heavy quark pairs, and to test theoretical calculations at higher orders. 

\begin{figure}[H]

\includegraphics[scale=0.33]{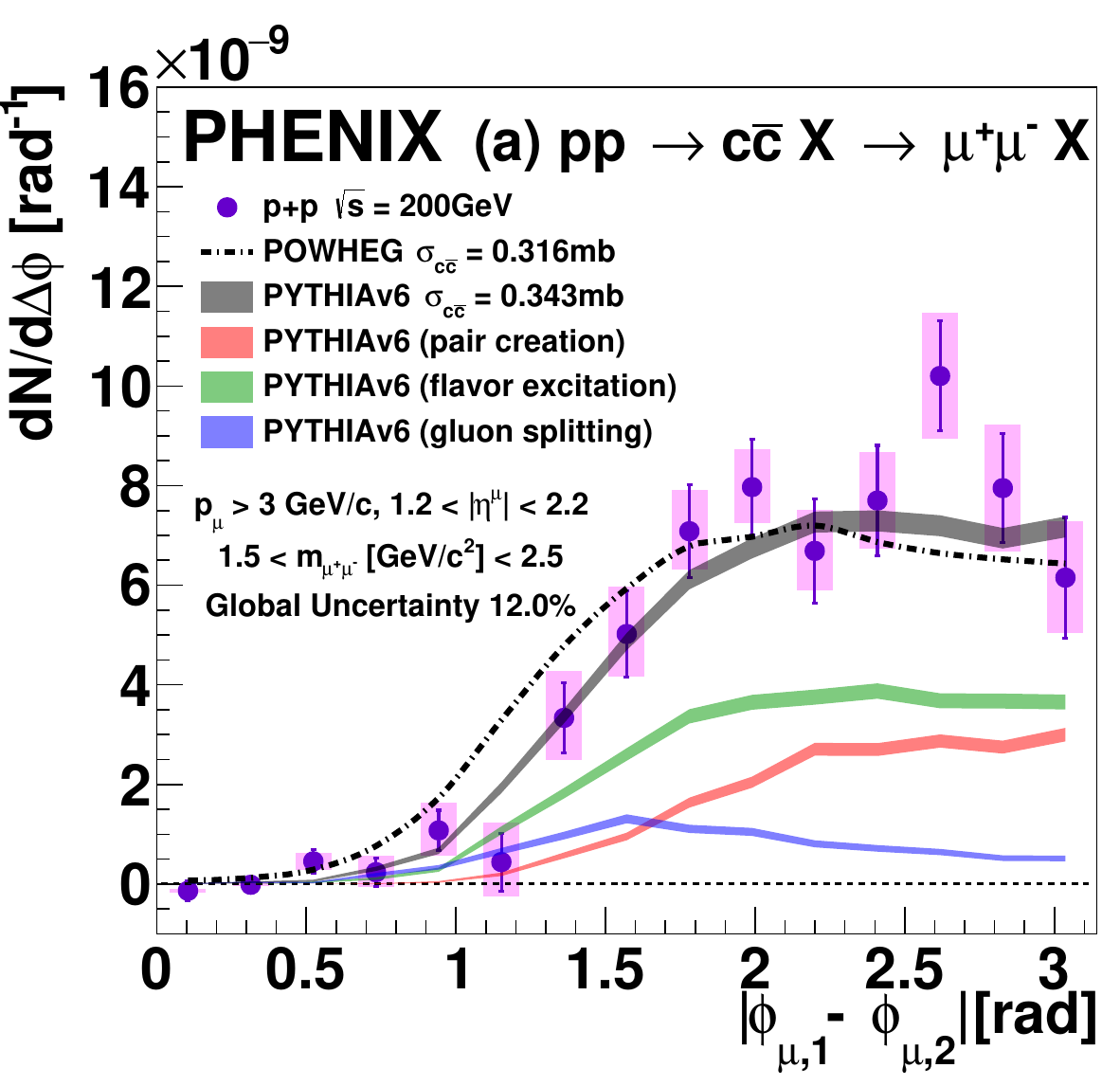}
\includegraphics[scale=0.33]{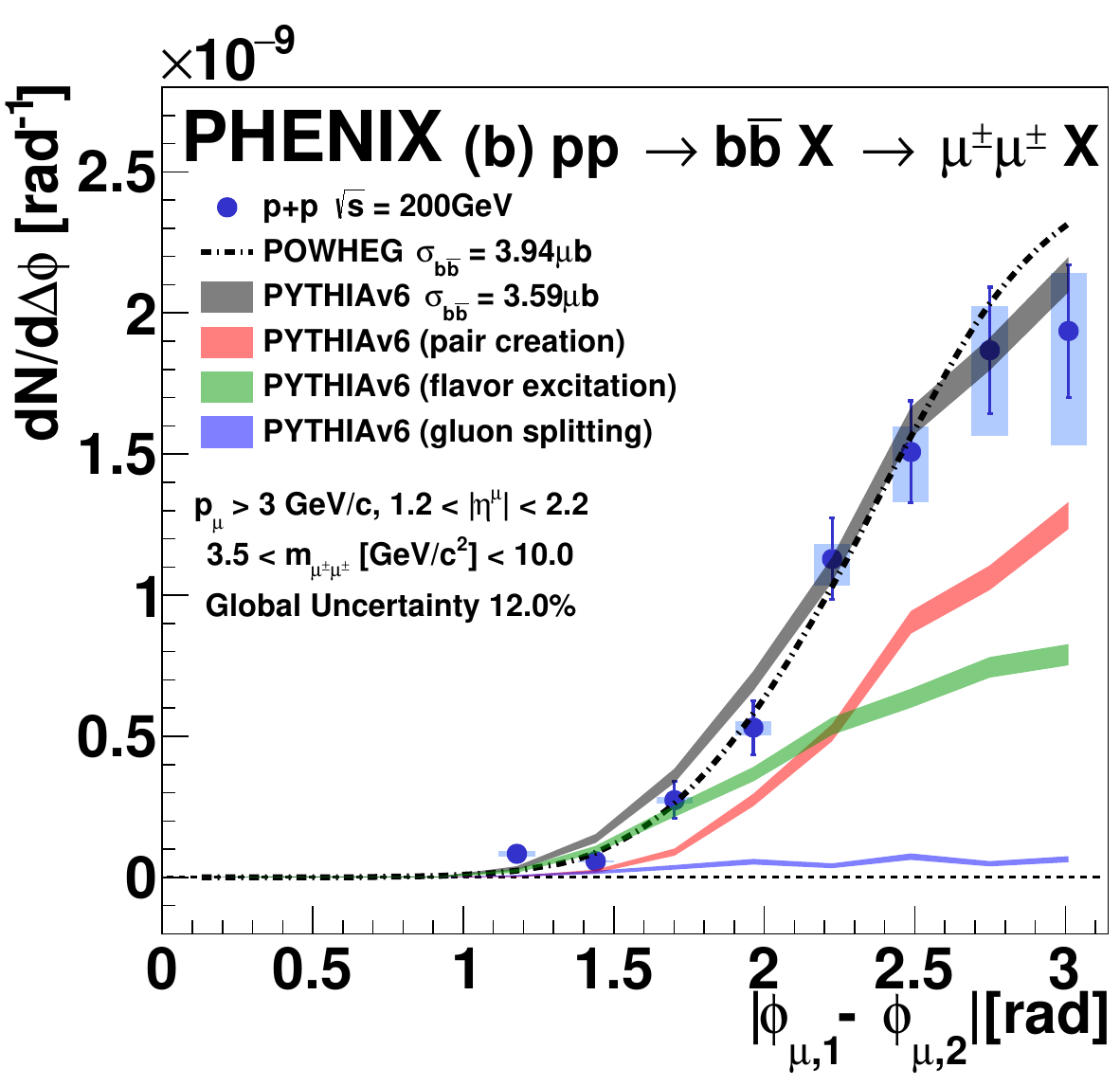}
\caption{The 
 azimuthal opening angle distribution of $\mu\mu$ pairs from $\rm{c}\bar{\rm{c}}$ (\textbf{a
}) and $\rm{b}\bar{\rm{b}}$ (\textbf{b
})  decays, measured by the PHENIX Collaboration in pp collisions at $\sqrt{s}=200$~GeV. The data are compared with the distributions calculated with POWHEG and PYTHIA event generators, where the different production mechanisms are also shown separately~\cite{PHENIX:2018dwt}.}
\label{fig:PhenixPpMuMuDPhi}
\end{figure}

The PHENIX Collaboration performed a study of azimuthal correlations of $\mu\mu$ pairs from heavy-flavor hadron decays in pp collisions at $\sqrt{s}= 200$ GeV at forward and backward rapidity $(1.2 < |\eta| < 2.2)$~\cite{PHENIX:2018dwt}. The $\mu\mu$ pairs have contributions from $\rm{c}\bar{\rm{c}}$, $\rm{b}\bar{\rm{b}}$, Drell--Yan, and hadronic pairs (kaons and pions), which are distinguished using template fits to opposite- and like-sign spectra in mass and transverse momentum, $p_{\rm{T}}$. While the decays from $\rm{c}\bar{\rm{c}}$ and Drell--Yan mechanisms contribute to opposite-sign pairs only ($\mu^{+}\mu^{-}$), decays from $\rm{b}\bar{\rm{b}}$ pairs populate the like-sign distribution ($\mu^{\pm}\mu^{\pm}$) as well, either because of combination of $\rm{B} \rightarrow$ $\mu$ and $\rm{B}  \rightarrow \rm{D} \rightarrow$$\mu$ decay chains or decays following $\rm{B}^{0} \bar{\rm{B}^{0}}$ oscillations. Different components contribute with different relative abundances to the muon pair continuum in different mass regions of $\mu^{+}\mu^{-}$ and $\mu^{\pm}\mu^{\pm}$ pairs. The contributions from $\rm{c}\bar{\rm{c}}$ and $\rm{b}\bar{\rm{b}}$ are separated considering mass regions where they dominate. This corresponds to $ 1.5 < m_{\mu^+\mu^-} < 2.5$~GeV$/c^2$ for charm, and $3.5 < m_{\mu^\pm \mu^\pm} < 10$ GeV$/c^2$ for beauty. The azimuthal opening angle distributions for $\mu\mu$ pairs from $\rm{c}\bar{\rm{c}}$ and $\rm{b}\bar{\rm{b}}$ decays, for muons with $p > 3$ GeV$/c$ and $1.2 < |\eta| < 2.2$, are shown in Figure~\ref{fig:PhenixPpMuMuDPhi}. The measurements are compared with model calculations based on PYTHIA~\cite{Sjostrand:2006za} and POWHEG~\cite{Frixione:2007nw}. The distribution from PYTHIA describes the data in the probed kinematic range for both $\rm{c}\bar{\rm{c}}$ and $\rm{b}\bar{\rm{b}}$. While the distribution from POWHEG simulations for $\rm{c}\bar{\rm{c}}$ are wider compared to the one from PYTHIA, they are more similar for $\rm{b}\bar{\rm{b}}$. Both PYTHIA and POWHEG use the PYTHIA fragmentation scheme and very similar parton distribution functions. The differences between the model calculation for charm--origin correlations could originate from the different underlying correlation between $\rm{c}$ and $\bar{\rm{c}}$ quarks emerging from the hard process. While POWHEG implements NLO calculations for evaluating the hard parton scattering matrix elements, PYTHIA evaluates them at LO and mimics the NLO processes in the parton shower.  
\vspace{-6pt}

\begin{figure}[H]

\includegraphics[scale=0.5]{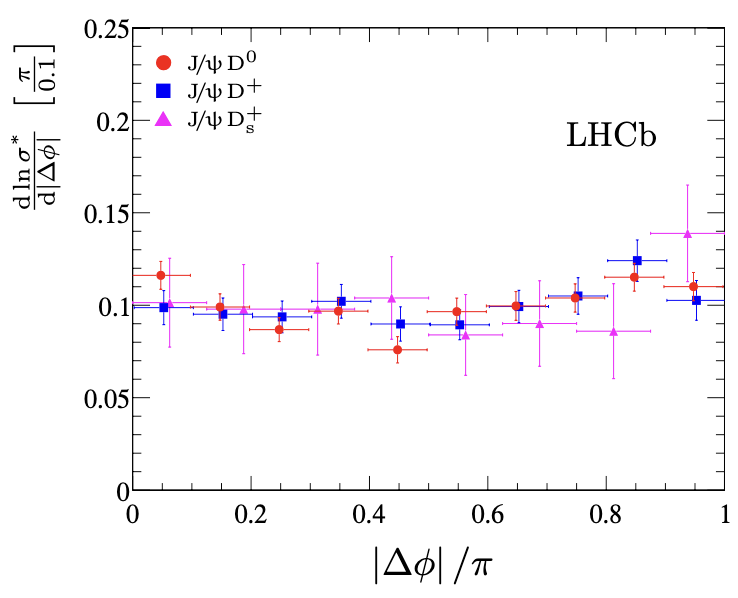}
\includegraphics[scale=0.5]{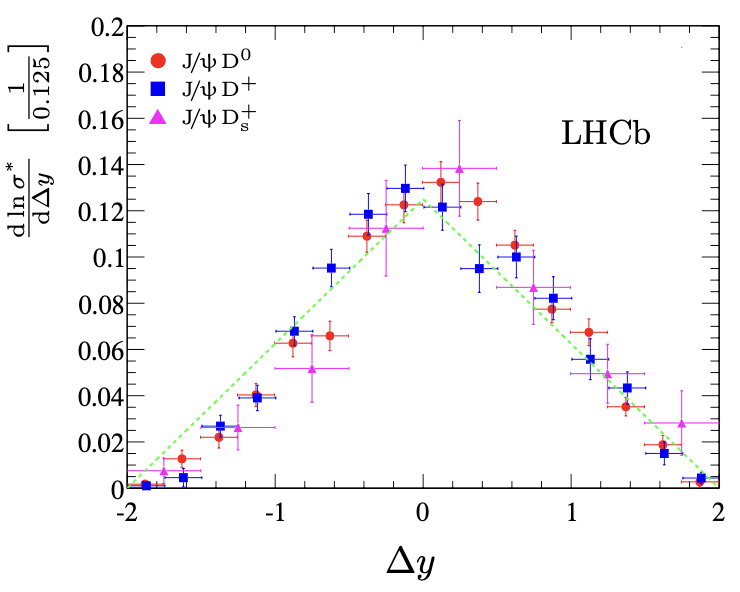}
\caption{Distributions 
 of the azimuthal angle (\textbf{left
}) and rapidity (\textbf{right
}) differences between $\rm{J}/$$\psi$ and different charm mesons ($C$) for $2 < y^{\rm{J}/}$$^{\psi}, y^C < 4$,  $p_{\rm{T}}$$^{,{\rm{J}/}}$$^{{\psi}} < 12$ GeV$/c$, and $3 < p_{\rm{T}}^C < 12$ GeV$/c$, in pp collisions at $\sqrt{s}=7$ TeV, measured by the LHCb Collaboration. The dashed line in the right plot shows the expected $\Delta y$ distribution for uncorrelated pairs~\cite{LHCb:2012aiv}.}
\label{fig:LHCbJPsiCCorr}
\end{figure}
\begin{figure}[H]

\includegraphics[scale=0.5]{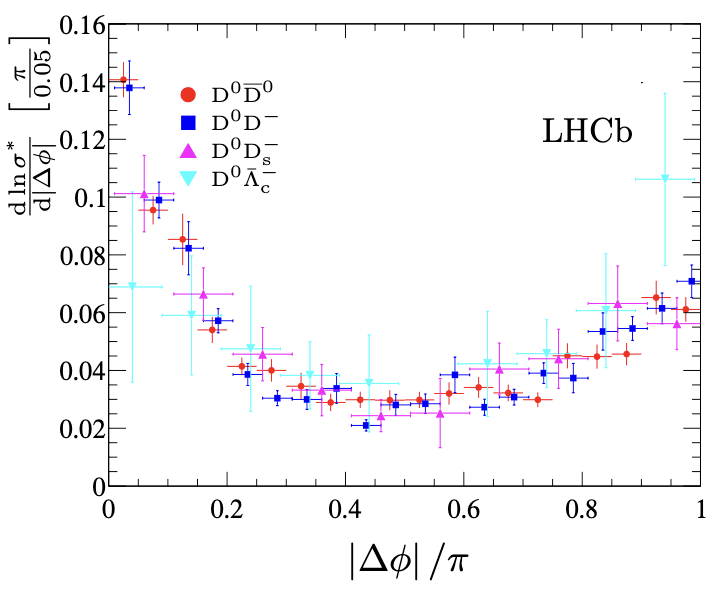}
\includegraphics[scale=0.52]{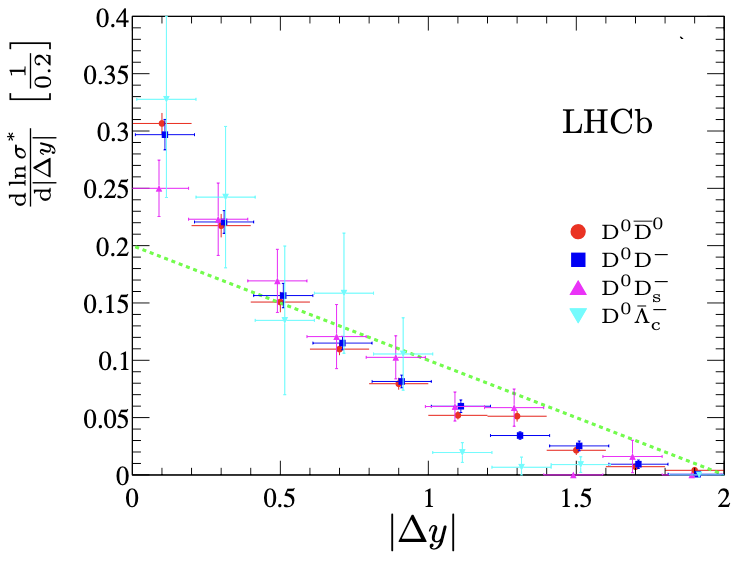}
\caption{Distributions 
 of the azimuthal angle (\textbf{left
}) and rapidity (\textbf{right
}) differences between charm and anti-charm mesons ($C\bar{C}$), for $2 < y^C < 4$ and $3 < p_{\rm{T}}^C < 12$ GeV$/c$ in pp collisions at $\sqrt{s}=7$ TeV, measured by the LHCb Collaboration. The dashed line in the right plot shows the expected $\Delta y$ distribution for uncorrelated pairs~\cite{LHCb:2012aiv}.}
\label{fig:LHCbCCbarCorr}
\end{figure}

To study the multiple production of charm states in a single pp collision, and understand contributions from double parton scattering (DPS)~\cite{Kom:2011bd,Baranov:2011ch,Novoselov:2011ff,Luszczak:2011zp}, the LHCb Collaboration performed measurements of J$/\psi$ production with an associated open-charm hadron (either $\rm{D}^{0}$, $\rm{D}^{+}$, $\rm{D}_s^+$ or $\Lambda_c^{+}$), and of double open-charm hadron production in pp collisions at \mbox{$\sqrt{s}$ = 7 TeV~\cite{LHCb:2012aiv}.} These measurements can probe the quarkonium production mechanism~\cite{Brodsky:2009cf} and contributions from the intrinsic charm content of the proton~\cite{Brodsky:1980pb} to the total cross-section. The J$/\psi$ and open-charm hadron production is denoted as J$/\psi C$ and the double open-charm hadron production as $CC$, with a control channel using $\rm{c}\bar{\rm{c}}$ events denoted as $C\bar{C}$. The measurements are performed in the LHCb fiducial region ($2 <y^{\rm{J}}$$^{/\psi} < 4.5$, $p_{\rm{T}}^{\rm{J}/}$$^{\psi} < 10$~GeV$/c$). The azimuthal angle and rapidity distributions between J$/\psi$ and charm hadrons in J$/\psi C$ events are shown in Figure~\ref{fig:LHCbJPsiCCorr}. No significant azimuthal correlation is observed and the $\Delta y$ distribution shows a triangular shape consistent with what is expected when the rapidity distribution for single-charm hadrons is flat and in the absence of physical correlations between the two particles. A similar trend in $\Delta\varphi$ and $\Delta y$ is observed for $CC$ pairs, with a slightly enhanced back-to-back configuration in $\Delta\varphi$, visible for $CC$ pairs compared to J$/\psi C$ pairs. The absence of significant azimuthal or rapidity correlations for J$/\psi C$ pairs could support the DPS hypothesis, but no comparison with model predictions are provided in the publication. In contrast, $C\bar{C}$ pairs show a clear enhancement in the $\Delta\varphi$ distribution at small $|\Delta\varphi|$, consistent with $\rm{c}\bar{\rm{c}}$ production via the gluon splitting mechanism at NLO~\cite{Norrbin:2000zc}, together with a significant contribution of back-to-back correlations, consistent with the topology of an LO production, as shown in Figure~\ref{fig:LHCbCCbarCorr}. This figure also shows a small enhancement at small $\Delta y$, which is also consistent with gluon splitting topologies, unlike the distributions for $CC$ and J$/\psi C$ pairs. 
\vspace{-6pt}
\vspace{-8pt}

\begin{figure}[H]

\includegraphics[scale=0.34]{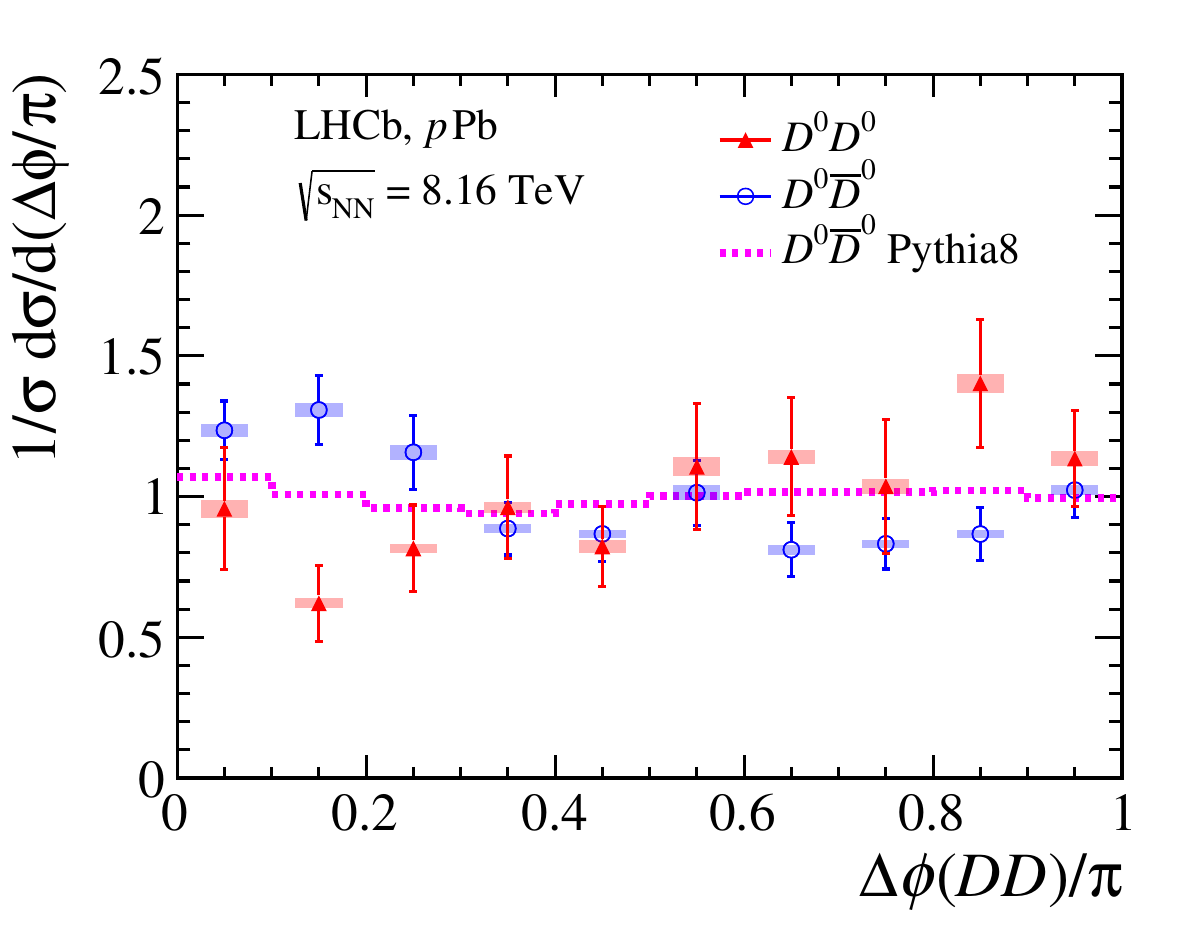}
\includegraphics[scale=0.34]{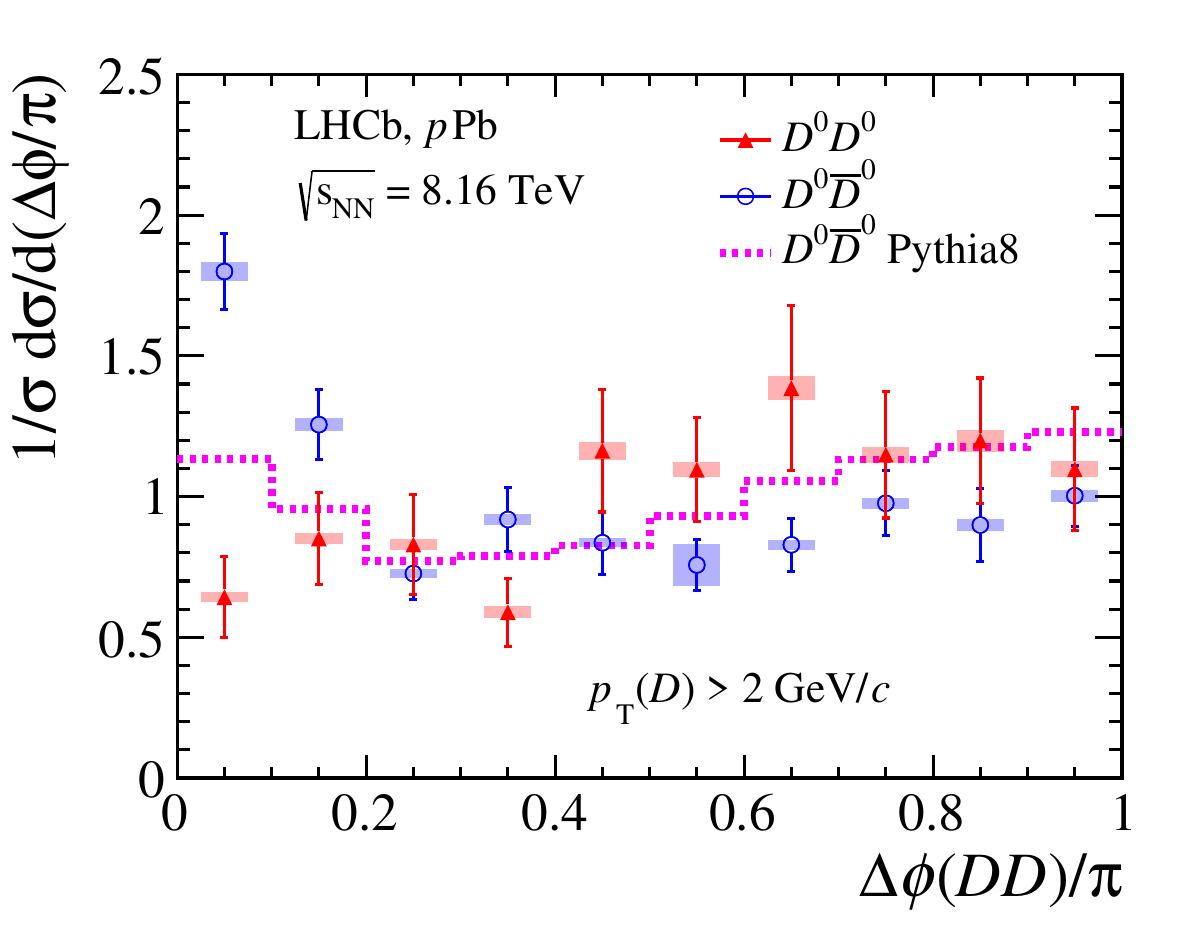}
\caption{The $\Delta\varphi$ distribution for $\rm{D}^0 \rm{D}^0$ (red) and $\rm{D}^0\bar{\rm{D}^0}$ (blue) pairs measured by the LHCb Collaboration in p--Pb collisions at $\sqrt{s_{\rm{NN}}} = 8.16$~TeV, compared with PYTHIA8 simulations (magenta dashed line), without (\textbf{left}) and with (\textbf{right}) a $p_{\rm{T}}^{\rm{D}^0} > 2$ GeV$/c$ requirement. Vertical bars (filled box) represent statistical (systematic) uncertainties~\cite{LHCb:2020jse}.}
\label{fig:LHCbCCbarCorr_pPb}
\end{figure}

The production of heavy quarks via single parton scattering (SPS) or via DPS can be affected by cold-nuclear-matter effects~\cite{Eskola:2009uj,deFlorian:2003qf,Hirai:2007sx, Fujii:2013yja,Tribedy:2011aa,Albacete:2012xq,Rezaeian:2012ye, Accardi:2009qv,Salgado:2011wc, Vogt:2018oje, Vogt:2019xmm} in proton--ion collisions. In particular, DPS production can be very sensitive to the nuclear PDF in p--A as it involves two parton pairs, including possible dependence on the position of the partons inside the nucleus~\cite{Shao:2020acd}. Based on the Glauber model~\cite{Miller:2007ri}, in the absence of nuclear effects, the SPS production cross-section is expected to scale with the ion mass number. DPS production, on the other hand, is enhanced compared to a mass number scaling due to collisions of partons from two different nucleons in the ion, and the enhancement factor is about three in proton--lead (p--Pb) collisions~\cite{Luszczak:2011zp, Cazaroto:2016nmu, Helenius:2019uge}. This feature was studied by the LHCb Collaboration by measuring the production of D meson pairs ($\rm{D}^0, \rm{D}^+, \rm{D}^+_s$ mesons), as well as J/$\psi \rm{D}$ meson pairs 
in p--Pb collisions at $\sqrt{s_{\rm{NN}}}=8.16$ TeV~\cite{LHCb:2020jse}. 
Like-sign (LS) pairs, where two hadrons have the same charm quark charge, and opposite-sign (OS) pairs, where they have opposite charm charge, are considered. Pairs of OS charm hadrons are dominantly produced from a $\rm{c}\bar{\rm{c}}$ pair via SPS; thus, the kinematics of the two hadrons are correlated, while DPS leads to both correlated and uncorrelated OS pairs. The kinematic correlation between the two charm hadrons was investigated using the two-charm invariant mass ($m_{\rm{DD}}$) and their relative azimuthal angle $\Delta\varphi$. The $\Delta\varphi$ distribution for LS $\rm{D}^0 \rm{D}^0$ pairs and OS $\rm{D}^0 \bar{\rm{D}^0}$ pairs for all $p_{\rm{T}}$ and when requiring a $p_{\rm{T}}^{\rm{D}^0} > 2$ GeV$/c$ condition is shown in Figure~\ref{fig:LHCbCCbarCorr_pPb}. Without the $p_{\rm{T}}$ condition, the $\Delta\varphi$ distribution is almost uniform for both LS and OS pairs, similar to that predicted by PYTHIA8 simulation. However, with the $p_{\rm{T}}^{\rm{D}^0} > 2$ GeV$/c$ condition, the $\rm{D}^0 \bar{\rm{D}^0}$ pair distribution shows an enhancement at $\Delta\varphi \sim 0$, while the $\rm{D}^0 \rm{D}^0$ distribution is consistent with being flat; both are inconsistent with the predictions from PYTHIA8 simulations. The flat $\Delta\varphi$ behavior for $\rm{D}^0 \rm{D}^0$ is qualitatively consistent with a large DPS contribution in LS pair production. The effective cross-section and  nuclear modification factor for LS charm hadron pairs were also measured and were reported to be compatible with the expected enhancement by a factor of 3 for a DPS-over-SPS production ratio from pp to p--Pb collisions~\cite{LHCb:2020jse}.
\vspace{-10pt}

\begin{figure}[H]

\includegraphics[scale=0.33]{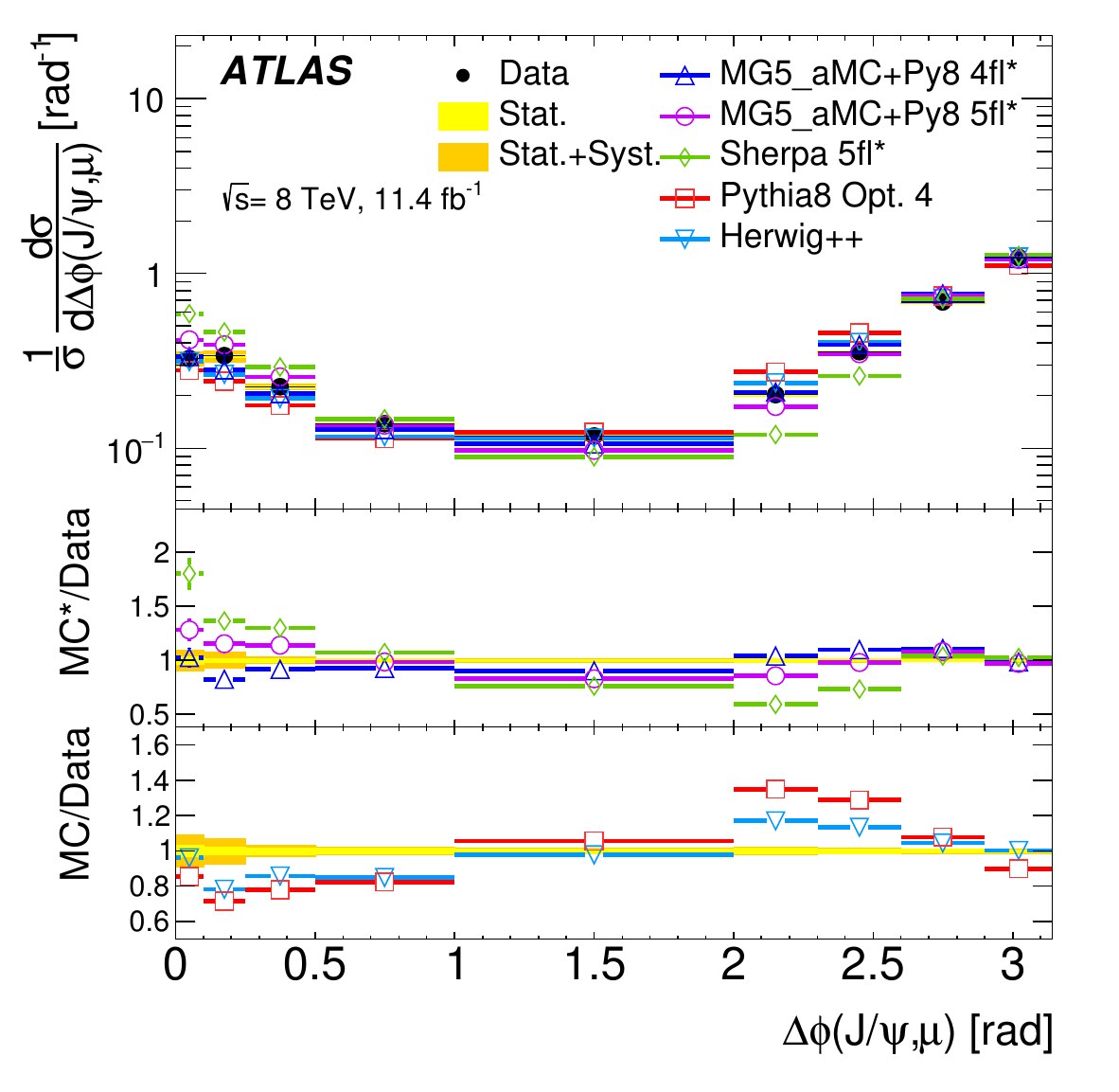}
\includegraphics[scale=0.33]{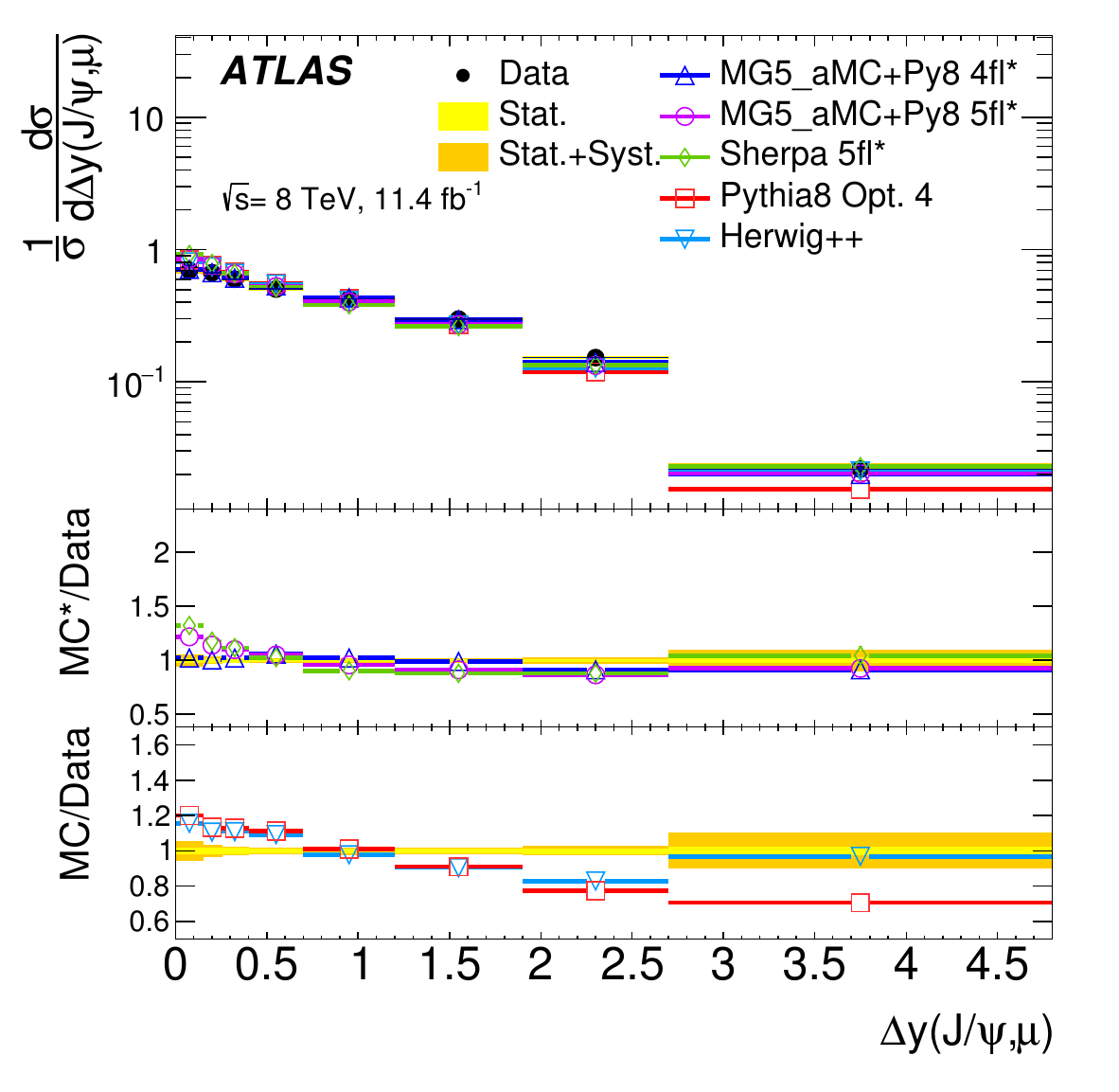}
\caption{Normalized 
 differential cross-section of J$/\psi$ and $\mu$ pairs as a function of $\Delta\varphi(\rm{J}$$/\psi,\mu)$ (\textbf{left}) and $\Delta y(J/\psi,\mu)$ (\textbf{right}) in pp collisions at $\sqrt{s}=8$~TeV, measured by the ATLAS Collaboration. Data are compared with predictions from PYTHIA8.2, HERWIG++, MADGRAPH5-AMC@NLO+PYTHIA8, and SHERPA Monte Carlo generators. The model-to-data ratios are shown in the bottom panels~\cite{ATLAS:2017wfq}.}
\label{fig:ATLASBBarCorr}
\end{figure}

The production of beauty quark pairs was studied by the ATLAS~\cite{ATLAS:2017wfq}, CMS~\cite{CMS:2011yuk}, and  LHCb~\cite{LHCb:2017bvf} Collaborations by measuring correlations of particles with beauty content. The ATLAS experiment measured the production of two {b-hadrons}, 
where one decays to J$/\psi(\rightarrow \mu\mu)$ + X and the other to $\mu$+Y, resulting in three muons in the final state in pp collisions at $\sqrt{s}=8$ TeV~\cite{ATLAS:2017wfq}. To probe the {b-hadron} production, several differential cross-sections were measured, such as $\Delta\varphi(\rm{J}$$/\psi,\mu)$, $\Delta y(\rm{J}$$/\psi,\mu)$, separation between the J$/\psi$ and the third muon in azimuth--rapidity plane $\Delta R (\rm{J}$$/\psi,\mu)$, mass of the three muon system $m(\rm{J}$$/\psi, \mu)$, etc. Since this review focuses on the angular correlation measurements, the $\Delta\varphi$ and $\Delta y$ distributions compared with predictions from PYTHIA8.2~\cite{Sjostrand:2007gs}, HERWIG++~\cite{Bahr:2008pv}, MADGRAPH5-AMC@NLO+PYTHIA8~\cite{Alwall:2014hca}, and SHERPA~\cite{Gleisberg:2008ta,Schumann:2007mg} Monte Carlo generators are shown in Figure~\ref{fig:ATLASBBarCorr}. The $\Delta\varphi$ prediction from HERWIG++ provides the best description of the data compared to others. The trends seen in $\Delta R (\rm{J}$$/\psi,\mu)$ are similar to $\Delta\varphi(\rm{J}$$/\psi,\mu)$. For $\Delta y(\rm{J}$$/\psi,\mu)$ distribution, the MADGRAPH5-AMC@NLO+PYTHIA8 and SHERPA predictions provide a good description of the data, while PYTHIA8 and HERWIG++ fail to describe data at high $\Delta y(\rm{J}$$/\psi,\mu)$. Different kinematic correlation observables can thus provide enhanced sensitivity to the underlying model differences and allow us to discriminate among them.  

\begin{figure}[H]

\includegraphics[scale=0.33]{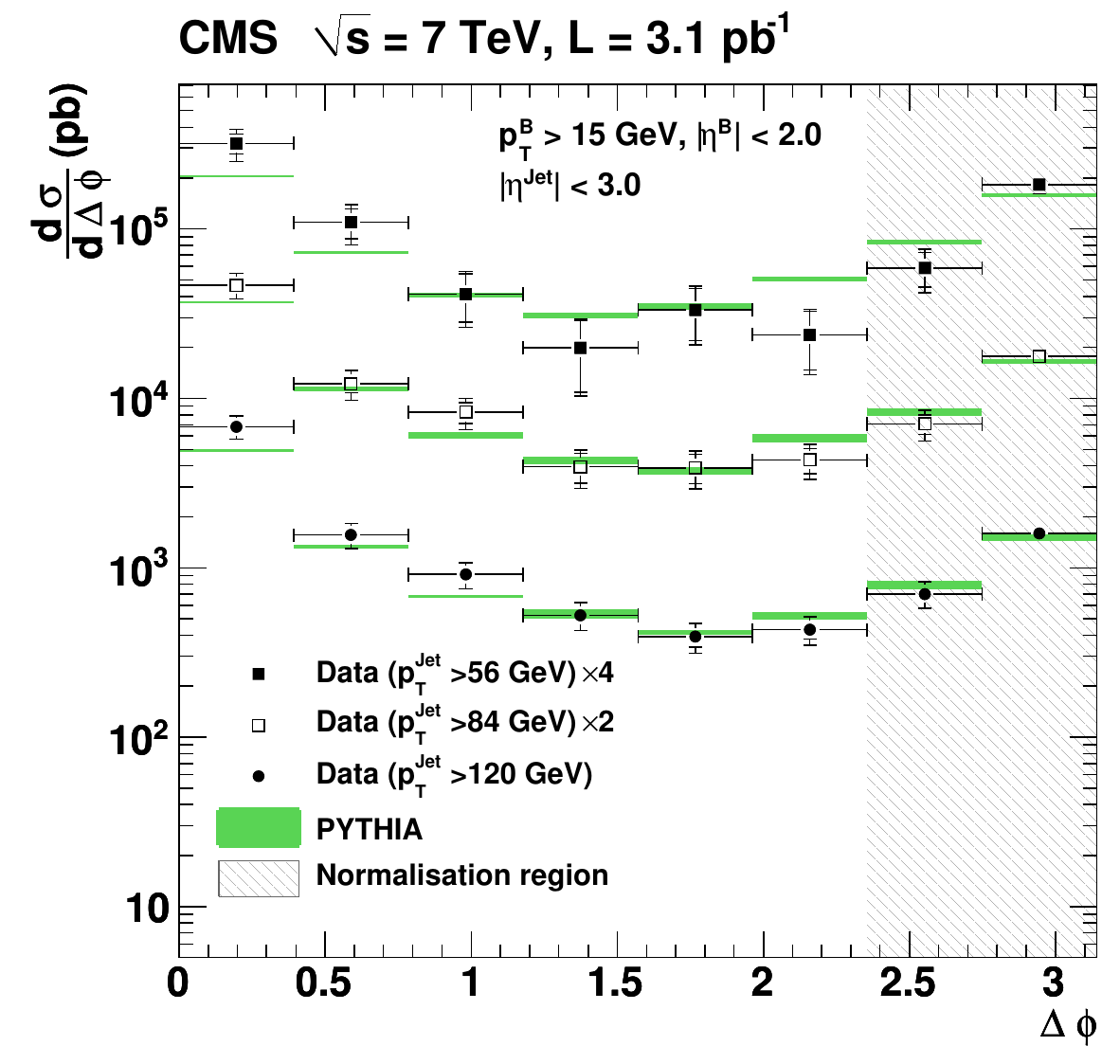}
\includegraphics[scale=0.33]{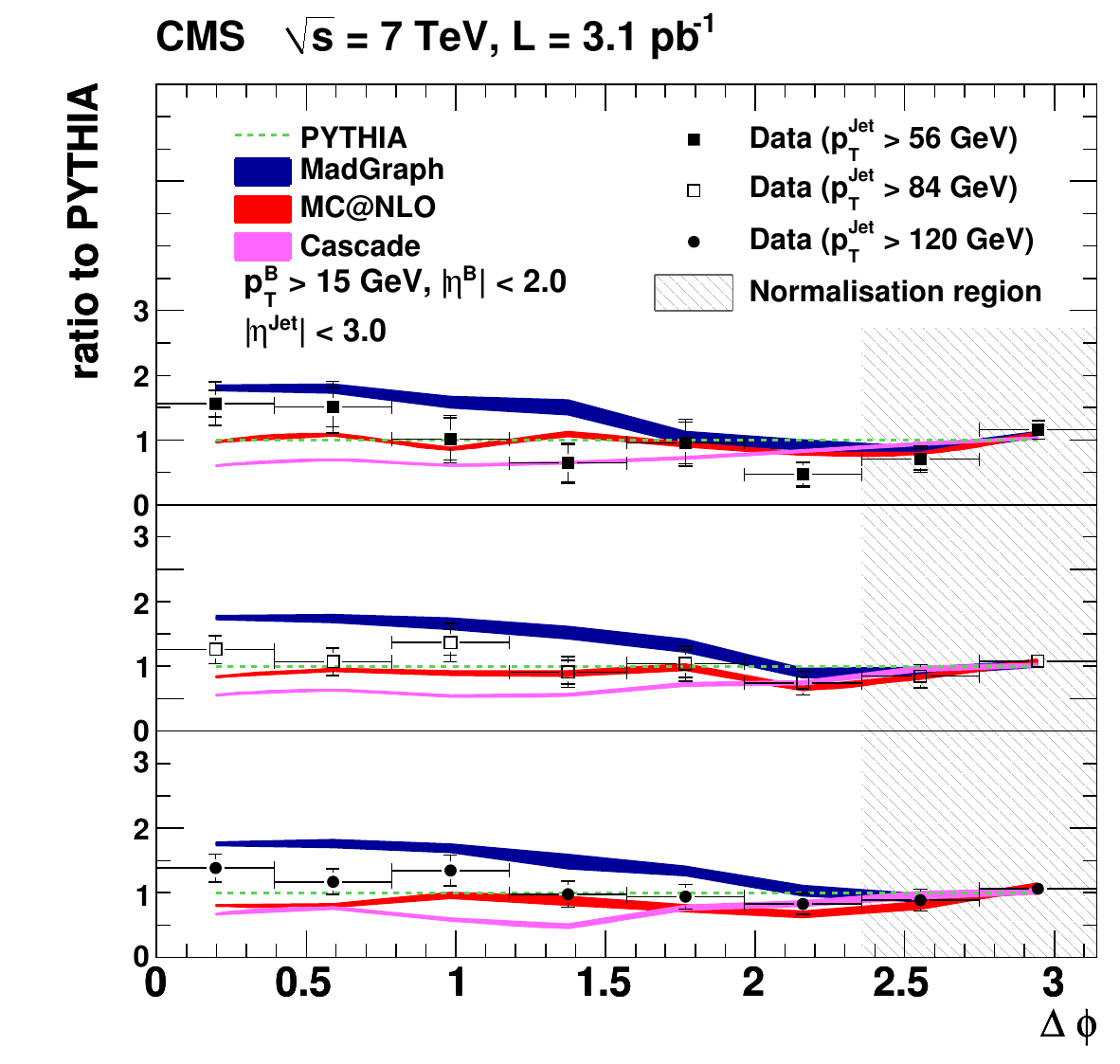}
\caption{(\textbf{Left}) Differential 
 $\rm{B}\bar{\rm{B}}$ production cross-section as a function of $\Delta\varphi$ for three leading jet $p_{\rm{T}}$ regions in pp collisions at $\sqrt{s}=7$~TeV, measured by the CMS Collaboration, and compared to PYTHIA predictions. For the data points, the error bars show the statistical (inner bars) and the total (outer bars) uncertainties. (\textbf{Right}) Ratio of the cross-section as a function of $\Delta\varphi$ for data, MADGRAPH, MC@NLO, and CASCADE models, with respect to PYTHIA predictions, for the three leading jet $p_{\rm{T}}$ regions. The simulations (shaded bands) are normalized to the region $\Delta\varphi > \frac{3}{4}\pi$. The widths of the shaded bands indicate the statistical uncertainties of the predictions~\cite{CMS:2011yuk}.}
\label{fig:CMSBBarCorr}
\end{figure}

\vspace{-8pt}
The azimuthal angular correlations between beauty hadron pairs, called here ``B hadrons'', were measured by the CMS Collaboration in pp collisions at $\sqrt{s}=7$ TeV, where the B hadrons were identified by the presence of displaced secondary vertices (SVs) from their decays~\cite{CMS:2011yuk}. The flight direction of the B hadron is defined by the direction connecting the primary and the secondary vertices. The angular correlation variables between B and $\bar{\rm{B}}$ hadrons, in events with two SVs, are calculated using their flight directions. 
While a back-to-back configuration is expected for LO processes, the B$\bar{\rm{B}}$ production at small opening angles directly relates to collinear emission processes at higher order (g $\rightarrow$ b$\bar{\rm{b}}$). The B$\bar{\rm{B}}$ pair production cross-section as a function of $\Delta\varphi$ for three different energy scales, characterized by the leading jet $p_{\rm{T}}$, is presented in Figure~\ref{fig:CMSBBarCorr} (left plot). A significantly large cross-section is observed at small angles, with values higher than at large angles, whose relative contribution increases with increasing jet $p_{\rm{T}}$. At higher energy scales, larger contributions from higher-order processes, for example, gluon radiation, are expected, resulting in more gluon splitting into B$\bar{\rm{B}}$ pairs. PYTHIA predictions~\cite{Sjostrand:2006za} are normalized to the region $\Delta\varphi > 3/4 \pi$, where the theoretical calculation is more reliable as the cross-section is expected to be dominated by the LO diagrams. Data and theory predictions from MADGRAPH~\cite{Maltoni:2002qb,Alwall:2007st}, MC@NLO~\cite{Frixione:2002ik,Frixione:2003ei,Frixione:2008ym}, and CASCADE~\cite{Jung:2000hk} models are compared, with respect to the PYTHIA prediction, as shown in Figure~\ref{fig:CMSBBarCorr} (right plot). It is observed that none of the predictions describe the data particularly well, in particular for the collinear region. The data lie between MADGRAPH and PYTHIA curves. 

\begin{figure}[H]

\includegraphics[scale=0.5]{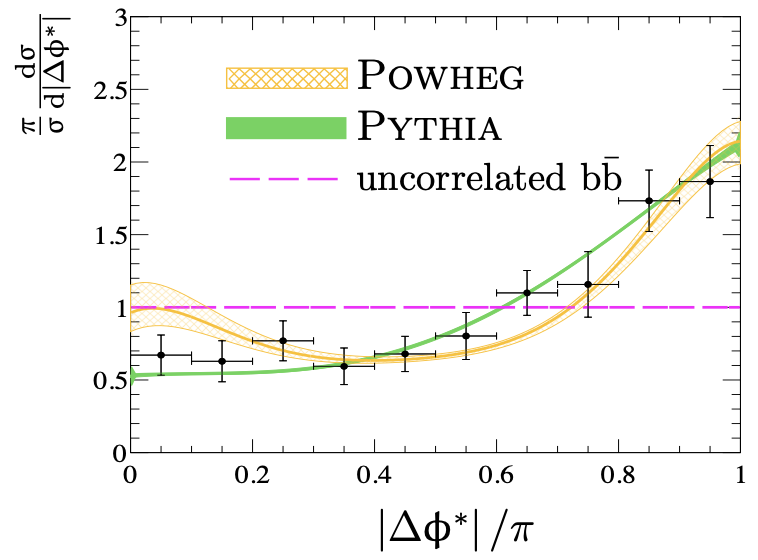}
\includegraphics[scale=0.5]{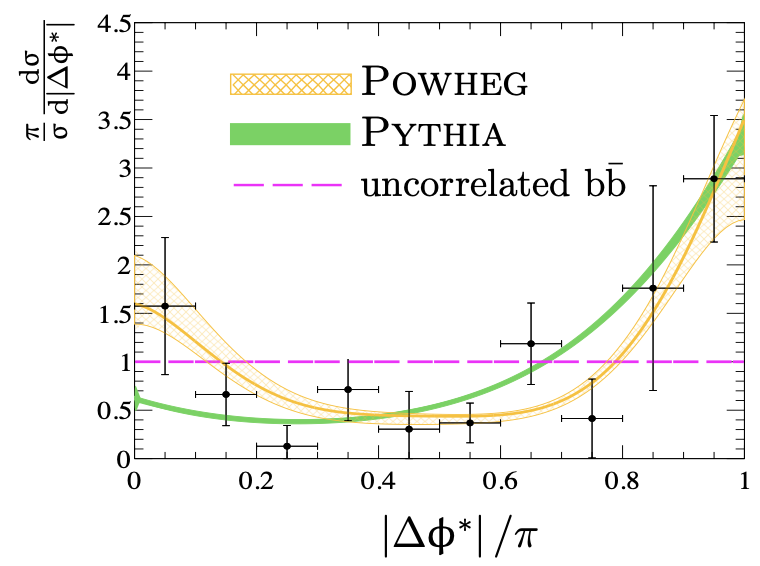}
\caption{Normalized 
 differential production cross-sections of B hadron pairs as a function of $|\Delta\varphi^*|$ for $p_{\rm{T}}^{\rm{J}}$$^{/\psi} > 3$ GeV/$c$ (\textbf{left
}) and $p_{\rm{T}}^{\rm{J}}$$^{/\psi} > 7$ GeV/$c$ (\textbf{right
}) in pp collisions at $\sqrt{s}=7$ and 8 TeV, measured by the LHCb Collaboration. The data are compared with POWHEG and PYTHIA predictions. The expectations for uncorrelated $\rm{b}\bar{\rm{b}}$ production are shown by the dashed magenta line. The uncertainties in the model predictions due to the choice of factorization and re-normalization scales are shown as solid bands~\cite{LHCb:2017bvf}.}
\label{fig:LHCbBBarCorr}
\end{figure}

The azimuthal and rapidity correlations in $\rm{b}\bar{\rm{b}}$ production in the forward rapidity region were investigated by LHCb experiment in pp collisions at $\sqrt{s}=7$ and 8 TeV by correlating pairs of beauty hadrons~\cite{LHCb:2017bvf}. The beauty hadrons were reconstructed via their inclusive decays into J$/\psi$ mesons (b$\rightarrow$J$/\psi X$). The $|\Delta\varphi^*|$ and $|\Delta\eta^*|$ variables, i.e., the difference in azimuthal angle $\varphi^*$ and pseudorapidity $\eta^*$ between the two beauty hadrons, estimated from the direction of the vector from the primary vertex to the decay vertex of the J$/\psi$ meson, were measured. The normalized differential production cross-sections as a function of $|\Delta\varphi^*|$ for $p_{\rm{T}}^{\rm{J}}$$^{/\psi} > 3$ and $ > 7$ GeV$/c$ are shown in Figure~\ref{fig:LHCbBBarCorr}. No significant enhancement is observed at small $|\Delta\varphi^*|$ at low $p_{\rm{T}}^{\rm{J}}$$^{/\psi}$, but a peak starts to appear in that region at higher $p_{\rm{T}}^{\rm{J}}$$^{/\psi}$, due to higher contribution from NLO processes. This observation is similar to the CMS measurement~\cite{CMS:2011yuk} from B hadron pairs. Compared to open-charm mesons~\cite{LHCb:2012aiv}, the small angle enhancement is observed at higher $p_{\rm{T}}$ since the contribution from processes like gluon splitting requires higher energy scales to produce higher-mass b quarks. The $|\Delta\varphi^*|$ distributions are compared with LO and NLO expectations from PYTHIA~\cite{Sjostrand:2007gs} and POWHEG~\cite{Gauld:2015yia} MC simulations, respectively. The prediction from an artificial data-driven model assuming uncorrelated $\rm{b}\bar{\rm{b}}$ production is also shown. At lower $p_{\rm{T}}^{\rm{J}}$$^{/\psi}$, the PYTHIA prediction describes the data well, suggesting that NLO effects in $\rm{b}\bar{\rm{b}}$ production in this kinematic region are small compared with the experimental precision. At higher $p_{\rm{T}}^{\rm{J}}$$^{/\psi}$, data are instead better described by POWHEG calculations. 
\vspace{-4pt}

\vspace{-6pt}
\section{Characterizing the Fragmentation of Heavy Quarks into Jets}
\label{sect:jetfragm}
In the absence of a surrounding nuclear medium, the process of hadronization of a heavy quark into colorless hadrons is generally described as a non-perturbative fragmentation of the quark into lower-momentum partons that, in the final state, are converted into colorless hadrons. At high $\pt$, this process produces a spray of particles traveling in a similar direction, called ``jet''.
A thorough characterization of the in-vacuum heavy quark fragmentation process can be obtained by measuring the angular correlation distribution between heavy-flavor ``trigger'' particles and other ``associated'' charged particles produced in the same event, in pp collisions. Under the assumption of a leading-order production process of the quark--anti quark pair, two peaks can be observed in the correlation distribution for $\Delta\varphi \approx 0$ (``near-side'' peak) and $\Delta\varphi \approx \pi$ (``away-side'' peak). These peaks are produced by particles in the jets emerging from the heavy quark pair fragmentation, which are approximately collinear with the quark's directions. 
While the presence of next-to-leading-order heavy quark production processes{---that are quite significant at LHC energies---}
breaks this picture for the away-side peak, the features of the near-side peak retains a clear connection to the original parton shower features.
In particular, studying the peak shape, its particle content and composition, and the $\pt$ distribution of its constituents for different kinematic regimes allows us to retrieve information about the heavy quark fragmentation process.
By comparing the features of the near-side peak with predictions from theoretical models or Monte Carlo simulations that implement different techniques to model the heavy quark fragmentation (and, in general, for the description of processes involving heavy quarks), it becomes possible to discriminate models and to validate those that yield the most accurate description of the data. In general, these comparisons allow us to determine constraints on the model configuration, parameters, and tuning.

In the presence of a deconfined medium, like the quark--gluon plasma produced in ultra-relativistic heavy-ion collisions, the hadronization process can be modified with respect to in-vacuum fragmentation. In particular, an additional hadronization mechanism, the coalescence, is expected to play a prominent role, where neighboring quarks in phase space recombine into higher-momentum bound hadrons~\cite{Fries:2003vb, Greco:2003mm, Ravagli:2007xx}. The role of coalescence, {as a competing mechanism} 
 to the quark fragmentation, is already hinted from studies of charm/hadron production ratios in heavy-ion collisions~\cite{STAR:2021tte, ALICE:2021bib, ALICE:2021kfc}. The modified hadronization should result in a significant modification of the final-state jet produced by the heavy quark. Such a modification can be evidenced and quantified by comparing the properties of the near-side peak of angular correlations between heavy-flavor particles and other particles, in Pb--Pb collisions to the reference system of pp collisions.

In this context, the ALICE Collaboration measured the azimuthal correlation distribution between D mesons and charged particles in pp collisions at $\s = 13$ TeV~\cite{ALICE:2021kpy}. A weighted average of the correlation distributions of $\Dzero$, $\Dplus$, and $\Dstar$ mesons was considered, at central rapidity ($|y| < 0.5$) in the transverse momentum range $3 < \pt^{\rm D} < 36$ GeV/$c$, while the associated particles were reconstructed in $\eta < 0.8$ for $\pt^{\rm assoc} > 0.3$ GeV/$c$. Only pairs with $|\deta|$ < 1 were considered. The correlation distribution was fitted with a function composed of a generalized Gaussian, describing the near-side peak, a Gaussian describing the away-side peak, and a constant, accounting for the physically uncorrelated pairs, assumed to be flat along $\dphi$. This model allowed for obtaining a quantitative description of the properties of the peaks in terms of their integral (peak yield) and width for the different kinematic ranges studied.
The correlation distributions and the near- and away-side peak properties were found to be consistent with the results obtained for lower center-of-mass energies ($\s = 5.02$ TeV~\cite{ALICE:2019oyn} and $\s = 7$ TeV~\cite{ALICE:2016clc}). The similarity of the near-side features implies that the charm quark fragmentation process is independent of the collision energy, at least for the energy ranges studied at the LHC.

Focusing further on the near-side correlation peak properties, a significant increase in its yield with increasing values of $\pt^{\rm D}$ was observed. This can be explained by the corresponding increase in charm quark $\pt$, on average, which implies that a larger amount of energy is available for the production of associated particles during its fragmentation. At the same time, a narrowing of the peak width can be observed when probing larger D meson $\pt$. Such an effect is related to the increased Lorentz boost of the charm quark, leading to a more collimated spray of particles produced by the fragmentation in the laboratory~frame.

\begin{figure}[H]
\includegraphics[width=\linewidth]{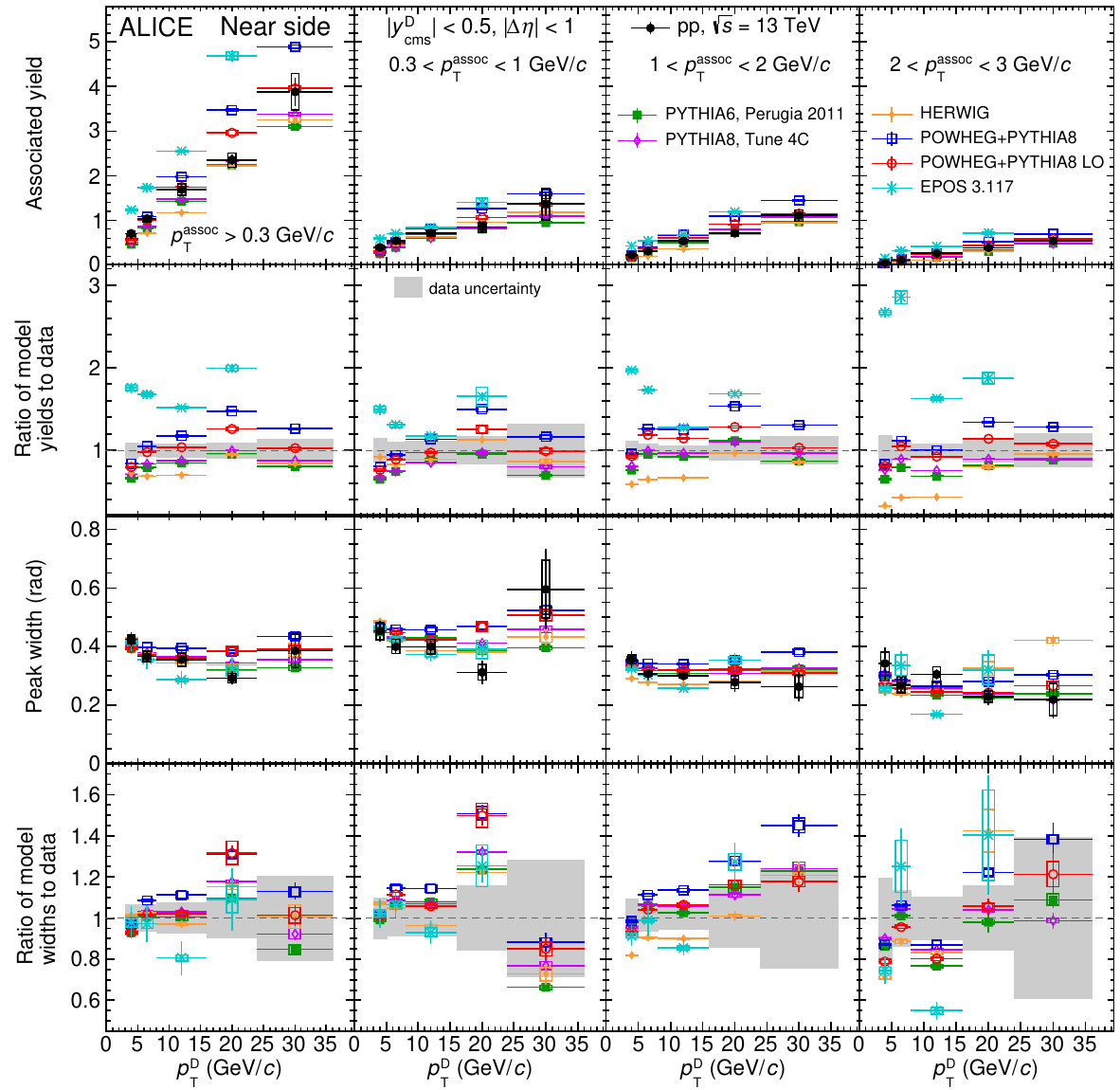}
\caption{\textbf{First 
} (\textbf{third}) \textbf{row}: near-side associated yields (widths) of azimuthal correlation distributions between D mesons and charged particles in pp collisions at $\s = 13$ TeV, measured by the ALICE Collaboration and predicted by several models, as a function of $\pt^{\rm D}$, for different $\pt^{\rm assoc}$ intervals. \textbf{Second} (\textbf{fourth}) \textbf{row}: model-to-data ratios of near-side associated yields (widths) of D meson and charged particle correlation distributions~\cite{ALICE:2021kpy}.
\label{Dh13TeV_ppvsModels_NS}}
\end{figure}

\textls[-25]{A comparison of the ALICE results with several model predictions, including PYTHIA8~\cite{Sjostrand:2007gs}} with 4C tune, POWHEG+PYTHIA8
~\cite{Nason:2004rx,Frixione:2007vw} using hard parton scattering matrix elements at LO or at NLO accuracy, HERWIG 7~\cite{Bellm:2015jjp}, and EPOS 3.117~\cite{Drescher:2000ha,Werner:2010aa}, is shown in Figure~\ref{Dh13TeV_ppvsModels_NS}. In particular, the near-side yields (widths) are shown in the first (third) row, and their model-to-data ratios are reported in the second (fourth) row.
Although all the models are able to reproduce the increase in peak yields for increasing D meson $\pt$, the strength of such a dependence, and the absolute values of the yields differ substantially among the various models. In particular, an ordering is found for the predicted near-side yields, with the lowest values observed for HERWIG (which tends to underestimate the data for \mbox{$\pt^{\rm D} < 16$ GeV/$c$),} followed by PYTHIA, POWHEG+PYTHIA, and EPOS, which overestimates the yield values in most of the $\pt$ intervals. Among the tested predictions, POWHEG+PYTHIA8 and PYTHIA generators are those that better reproduce the measured data and are thus more suited to quantify the number of particles emerging from charm quark fragmentation in association with the D meson. The various model predictions for the near-side widths show instead similar values in all the studied kinematic ranges, and overall, all models are consistent with the ALICE measurements within the uncertainties.

\begin{figure}[H]
\includegraphics[width=\linewidth]{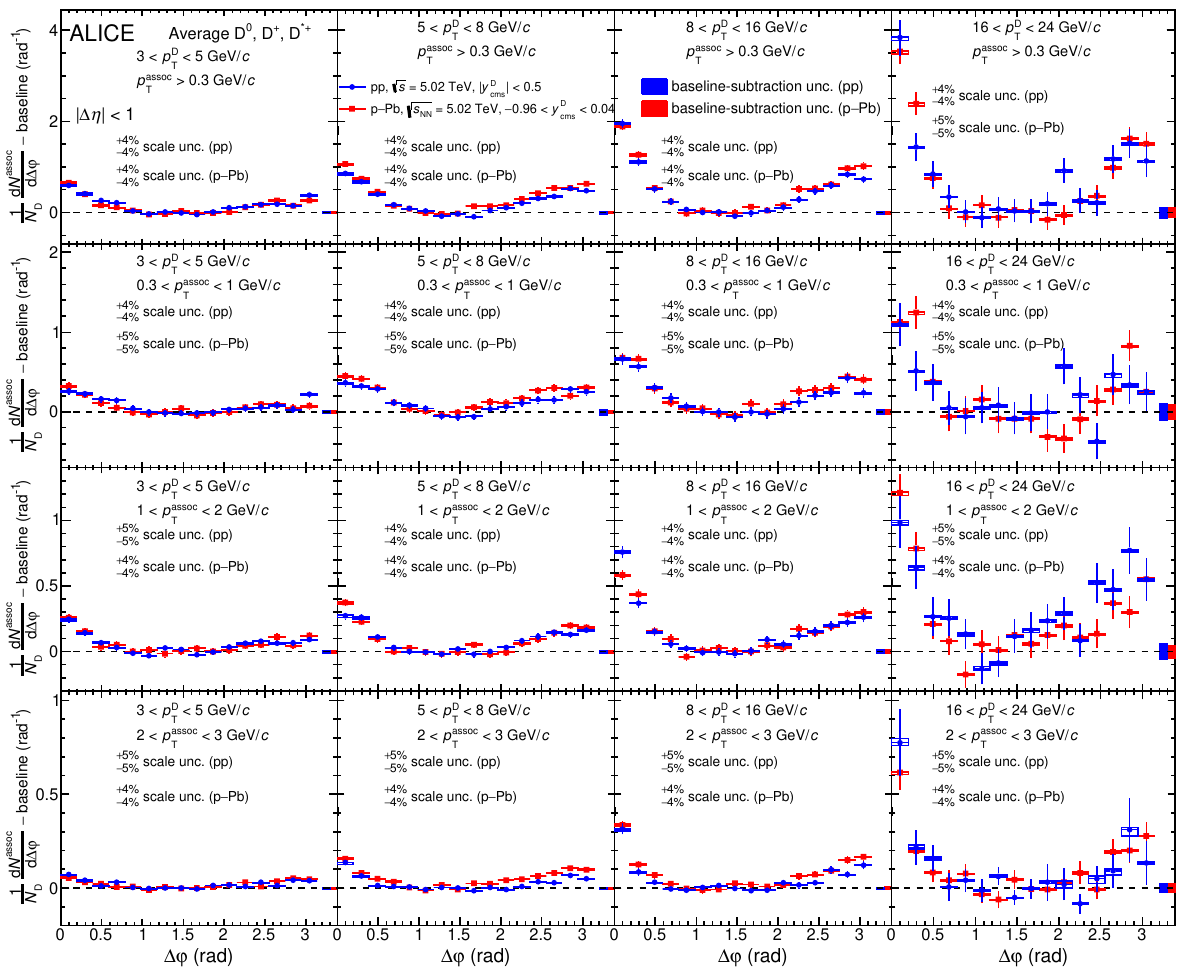}
\caption{Comparison 
 of the azimuthal correlation distributions between D mesons and charged particles in pp and p--Pb collisions at $\sNN = 5.02$ TeV, for the kinematic ranges studied by the ALICE Collaboration~\cite{ALICE:2019oyn}.
\label{DhpPb_ppvspPb_dPhi}}
\end{figure}

\begin{figure}[H]
\includegraphics[width=\linewidth]{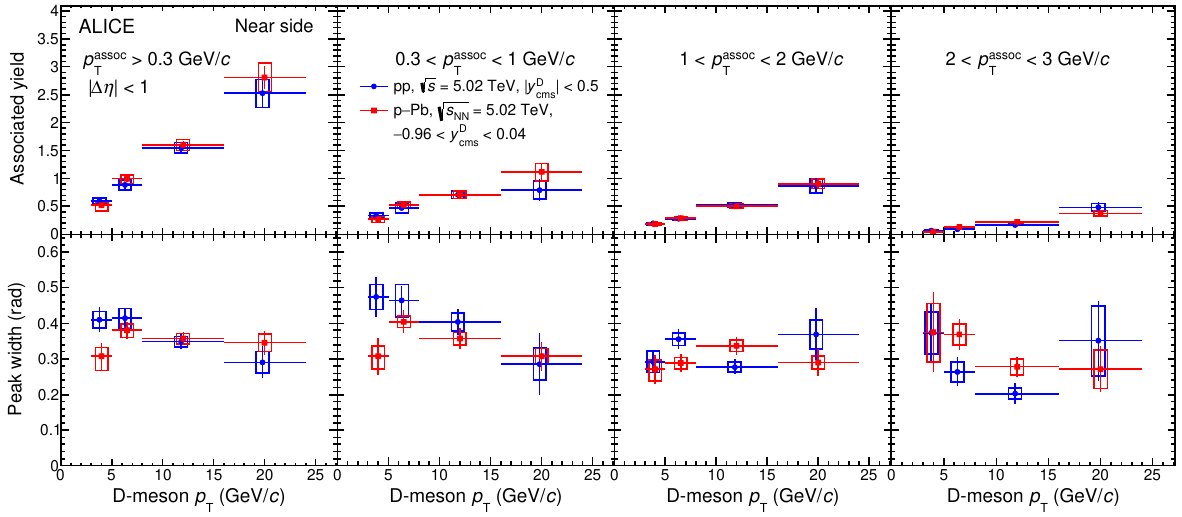}
\caption{Comparison 
 of near-side associated yields (\textbf{first row}) and widths (\textbf{second row}) of D meson and charged-particle azimuthal correlation distributions in pp and p--Pb collisions at $\sNN = 5.02$~TeV, measured by the ALICE Collaboration, as a function of $\pt^{\rm D}$, for different $\pt^{\rm assoc}$ intervals~\cite{ALICE:2019oyn}.
\label{DhpPb_ppvspPb_NS}}
\end{figure}

The in-vacuum behavior of heavy quarks, including their parton shower and hadronization, can be altered in the presence of a nuclear medium.
Measurements in p--Pb collisions are sensitive to the influence of cold-nuclear-matter effects on the heavy quarks, and can thus act as a reference to help disentangle and understand those modifications that are instead induced by the quark--gluon plasma environment in heavy-ion collisions.
To probe whether cold-nuclear-matter effects play a role in the charm quark fragmentation process, the ALICE Collaboration has measured the azimuthal correlation distribution of D mesons and charged particles at mid-rapidity in multiplicity-integrated p--Pb collisions at $\sNN = 5.02$ TeV. The same analysis technique used in the pp collision results discussed above~\cite{ALICE:2021kpy} was exploited, with the same kinematic coverage. The shape of the correlation distribution and its evolution with transverse momenta of D mesons and associated charged particles were compared, and found to be fully consistent, with those obtained in pp collisions at the same center-of-mass energy, as shown in Figure~\ref{DhpPb_ppvspPb_dPhi}.
Specifically, the comparison of the near-side peak yields and widths in the two collision systems is shown in Figure~\ref{DhpPb_ppvspPb_NS}. No modification of the near-side yield values, and the same increase in D meson $\pt$ measured in pp collisions were observed. For the near-side widths, the tendency for a collimation of the peak at larger $\pt^{\rm D}$ is possibly less pronounced, but pp and p--Pb results are similar within uncertainties.
From these results, no indications for a modified fragmentation process of charm quarks due to cold-nuclear-matter effects, or for any alteration of the charm hadronization mechanism, are observed.

In recent years, the high-energy nuclear physics community's interest in the dynamics of partons produced in small collision systems at very high multiplicities has grown. Although no clear modifications of high-$\pt$ particle production yields (beyond what is expected from nuclear modification of the parton distribution functions) have been measured, several indications of collective-like effects have been observed at the LHC in the recent past, including measurements in the heavy-flavor sector~\cite{ALICE:2018gyx,ATLAS:2019xqc,CMS:2018loe,CMS:2020qul} as discussed in more detail in Section~\ref{sect:collectivity}.
In general, the evaluation of flow coefficients in small collision systems is based on two- or multi-particle correlation techniques, and relies on the assumption that the contribution of jet peaks to the correlation distribution has negligible dependence on the event multiplicity, and can be removed from the high-multiplicity correlation distribution by measuring it in low-multiplicity collisions, where no collective effects are present.
Such an assumption can be tested by studying the jet fragmentation properties at different event multiplicities in small collision systems. In this regard, the ALICE experiment has studied the dependence of the azimuthal correlation distribution of D mesons and charged particles, and of its near-side peak features on the event multiplicity in pp collisions at \mbox{$\s = 13$ TeV~\cite{ALICE:2021kpy}} and p--Pb collisions at $\sNN = 5.02$ TeV~\cite{ALICE:2019oyn}. The results show that, within the experimental uncertainties, the near-side peak yields and widths are consistent for all the multiplicity ranges studied, suggesting a similar fragmentation of charm quark into final-state D mesons and other associated particles that is independent of the surrounding event activity.

\vspace{-12pt}

As an alternate approach for investigating the in-vacuum heavy quark fragmentation and possible impact of cold-nuclear-matter effects, the ALICE Collaboration has measured angular correlations between electrons produced from decays of heavy-flavor hadrons (called herein ``heavy-flavor decay electrons'' for simplicity) and charged particles in pp and p--Pb collisions at $\sNN = 5.02$ TeV~\cite{ALICE:2023kjg}. Compared to correlations with D mesons as heavy-flavor trigger particles, heavy-flavor decay electrons have a looser connection with the direction and momentum of the original heavy quark because of the hadron decay kinematics. On the other hand, such an analysis profits from a larger statistical sample, which allows for the transverse momentum range of the associated charged particles to be significantly extended (up to 7 GeV/$c$). Additionally, at high $\pt$, the sample of heavy-flavor decay electrons is dominated by those from beauty quarks, which enables the study of beauty quark fragmentation when focusing in the $\pt$ region above 7 GeV/$c$.
The study is performed at mid-rapidity, in the electron $\pt$ range $4 < \pt < 16$~GeV/$c$, considering pairs with pseudorapidity displacement $|\deta| < 1$. Also, in this case, a quantitative assessment of the quark-into-jet fragmentation is performed by fitting the correlation distribution with a function, composed as the sum of two von Mises functions, to model the near- and away-side peaks, plus a constant term.

\begin{figure}[H]

\includegraphics[scale=0.55]{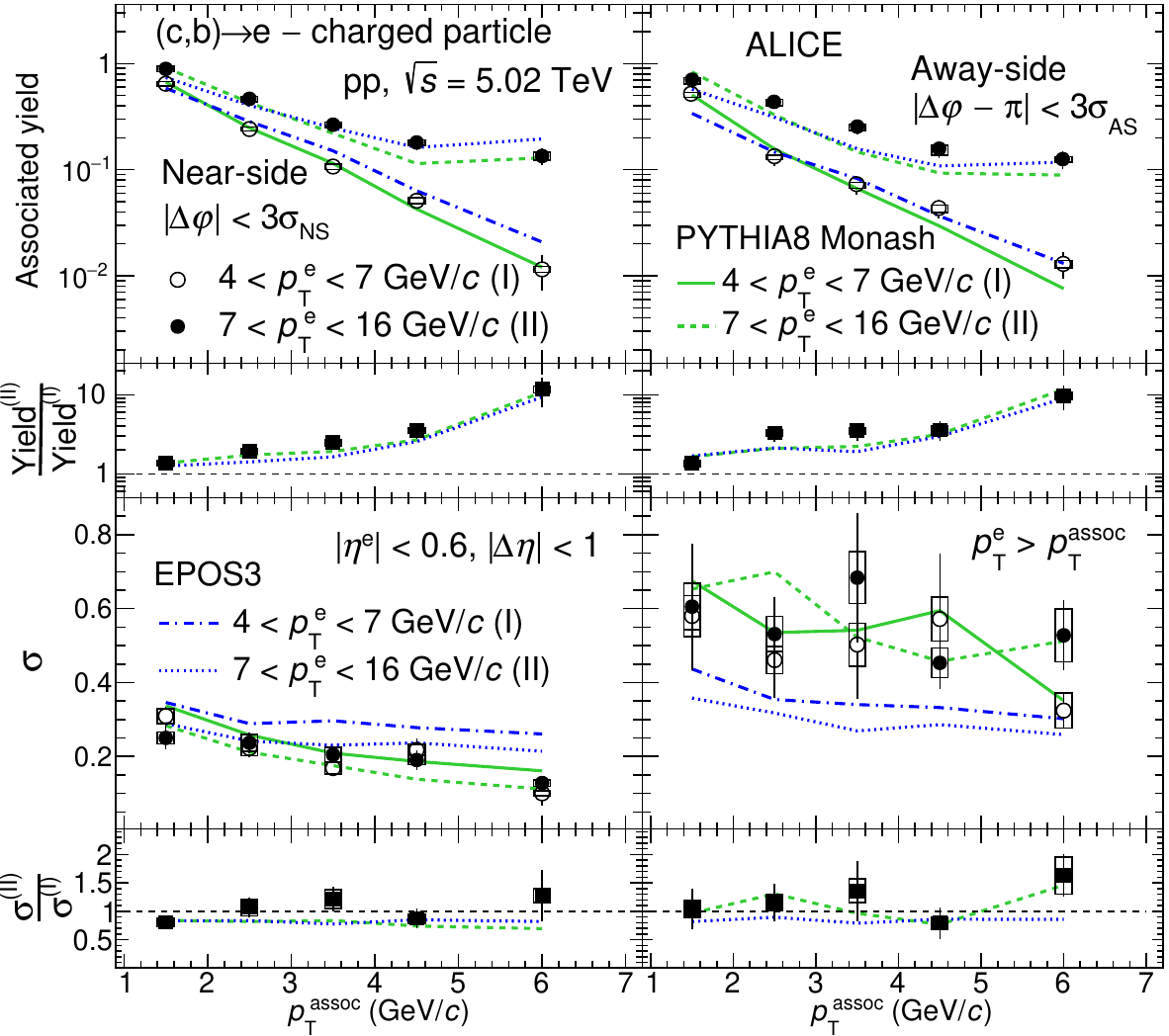}
\caption{Near- 
(\textbf{left}) and away-side (\textbf{right}) associated peak yields (\textbf{top}) and widths (\textbf{bottom}) of correlation distributions between heavy-flavor hadron decay electrons and charged particles in pp collisions at $\sNN = 5.02$, measured by the ALICE Collaboration and predicted by PYTHIA8 and EPOS3. The results are reported for two $\pt$ ranges, and the insets show the ratios of the observable distributions in higher to lower $\pt$ ranges considered~\cite{ALICE:2023kjg}.
\label{eh_ppvsModel_2ptBins}}
\end{figure}

In Figure~\ref{eh_ppvsModel_2ptBins}, the near-side peak yields and widths measured by ALICE in pp collisions are compared with predictions from the PYTHIA8 event generator~\cite{Sjostrand:2007gs} with the Monash tune and from the EPOS3 event generator~\cite{Drescher:2000ha,Werner:2010aa}.
Two transverse momentum ranges are considered for the electrons, i.e., $4< \pt^{\rm e} < 7$ GeV/$c$, with a balanced contribution between charm and beauty origins, and $7 < \pt^{\rm e} < 16$ GeV/$c$, where the large majority of electrons are produced by beauty hadron decays.
For both $\pt^{\rm e}$ ranges, the largest contribution of charged particles produced in the fragmentation is present below 2 GeV/$c$, pointing to a dominance of soft particle production from the quark fragmentation. The fraction of high-$\pt$-associated particles significantly increases when probing the high-$\pt^{\rm e}$ range, despite remaining subdominant. In addition, the absolute value of the yields is substantially larger in the $7 < \pt^{\rm e} < 16$ GeV/$c$ range compared to the $4< \pt^{\rm e} < 7$ GeV/$c$ interval. This is similar to what was observed for the D meson correlation with charged particles, and can be ascribed to the larger average energy of heavy quarks producing higher-$\pt$ electrons, which generally leads to an increased multiplicity of fragmenting particles.
The values of the near-side widths are fully consistent between the two $\pt^{\rm e}$ ranges, and point towards an emission of harder particles more collinear with the electron, while softer particles are emitted at larger angles.
Both PYTHIA8 and EPOS3 generators can successfully describe the near-side yield values, with EPOS3 predicting larger values for high $\pt$ of the associated tracks. While PYTHIA8 also correctly reproduces the near-side widths, EPOS3 tends to overestimate them at high $\pt^{\rm assoc}$, predicting a flatter trend than what is observed in the data.
The away-side peak yields and widths are also shown in the same figure. The away-side peak has a connection to the fragmentation of the heavy quark that did not lead to the production of the trigger particle, though such a connection is less direct than that of the near-side peak. In contrast, the away-side peak features are sensitive to the production mechanisms of the heavy quark pairs, which induce different angular topologies, as discussed in Section~\ref{sec:ProdMechPp}.
Very similar considerations as for the near-side can be drawn for the away-side peak yield values. The away-side peak widths are  about a factor 2 larger than those measured for the near-side, with significantly larger uncertainties. This is mainly due to the NLO production processes of heavy quarks, which break the back-to-back topology of the quark pairs, and to the additional smearing with respect to the original quark--anti-quark correlation distribution induced by the hadronic decays.

\begin{figure}[H]

\includegraphics[scale=0.55]{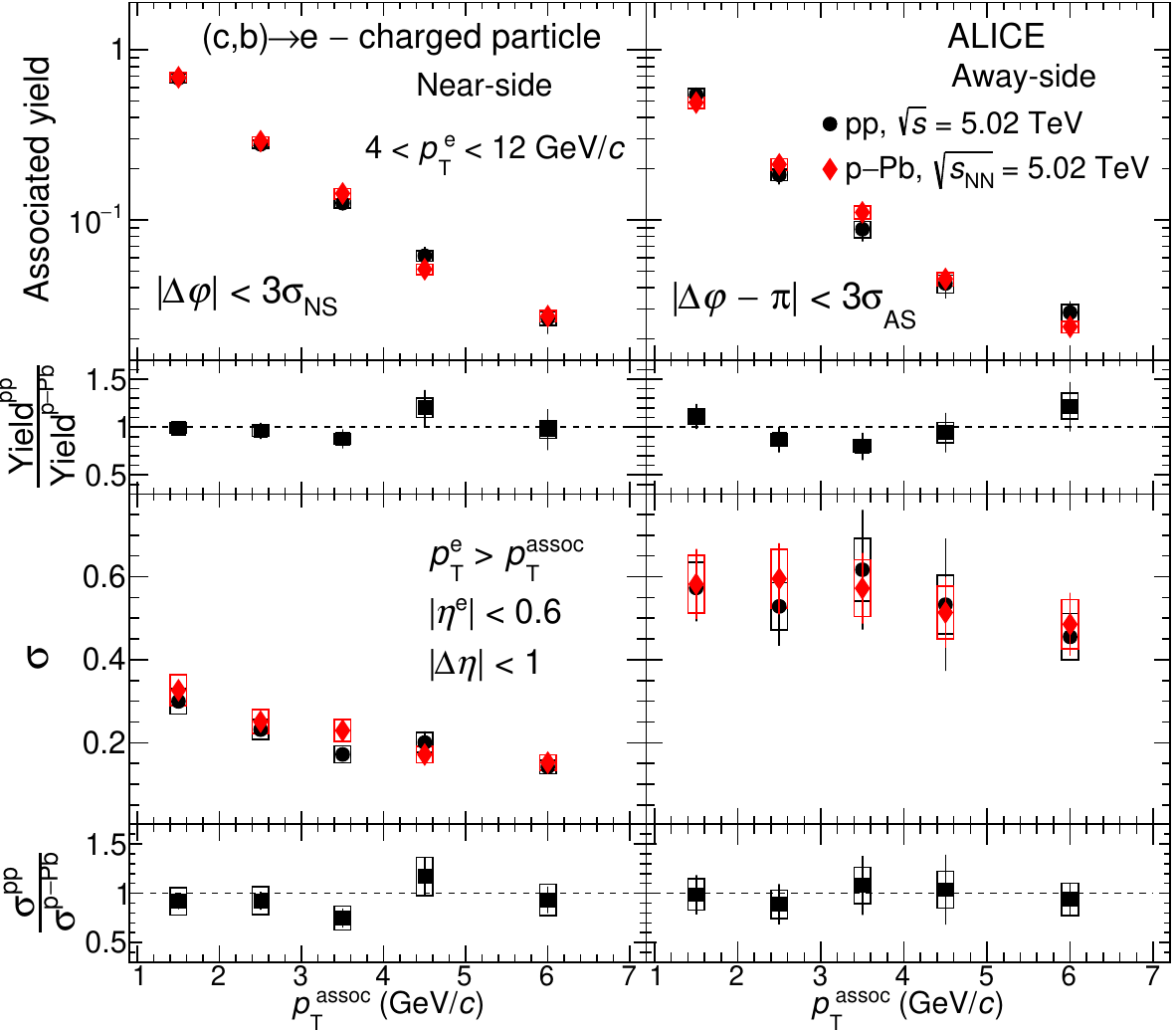}
\caption{Comparison 
 of near- (\textbf{left}) and away-side (\textbf{right})-associated peak yields (\textbf{top}) and widths (\textbf{bottom}) from the azimuthal correlation distributions of heavy-flavor hadron decay electrons and charged particles in pp and p--Pb collisions at $\sNN = 5.02$ TeV, measured by the ALICE Collaboration. The insets show the ratios of the observable distribution in pp to p--Pb collisions~\cite{ALICE:2023kjg}.
\label{eh_ppvspPb}}
\end{figure}

In the same publication, the near- and away-side peak properties of the azimuthal correlation between heavy-flavor decay electrons and charged particles in pp and p--Pb collisions at $\sNN = 5.02$ TeV are compared for $4 < \pt^{\rm e} < 12$ GeV/$c$ and various charged-particle transverse momentum intervals.
From the comparison, shown in Figure~\ref{eh_ppvspPb}, fully compatible peak yields and widths are found in the two collision systems. This observation holds also for the high $\pt^{\rm assoc}$ intervals not covered by previous D-hadron and charged particle correlation measurements~\cite{ALICE:2019oyn}. 
These results thus complement the findings observed for that analysis, and confirm that the charm quark fragmentation is unaffected by the presence of cold-nuclear-matter effects. No strong conclusions can be drawn for the beauty, given the lack of a specific comparison in a high-$\pt^{\rm e}$ interval.

\vspace{-22pt}

\section{Heavy Quark Energy Loss and Redistribution}
~\label{sect:corrHIC}
\vspace{-2pt}
In the presence of the quark--gluon plasma, produced in ultra-relativistic heavy-ion collisions, high-$p_{\rm{T}}$ partons produced in hard scatterings with high $Q^2$ lose energy via medium-induced gluon radiation and collisions with the medium constituents\mbox{~\cite{ALICE:2021rxa,ALICE:2021kfc,Gyulassy:1990ye,Baier:1996sk,Thoma:1990fm,Braaten:1991we}.} The processes by which quarks and gluons lose energy in the QGP can modify the internal structure and energy of the jet they produce, a phenomenon known as jet quenching. Through the passage of a jet, the QGP can itself be modified, due to the injection of energy and momentum lost by the jet into the plasma. Because of momentum conservation, a ``wake'' is induced in the medium as the jet loses energy and momentum, giving the medium a net momentum in the jet direction, yielding a correlation between the bulk dynamics of the medium and the jet direction~\cite{Casalderrey-Solana:2020rsj}.

Measurements of heavy-flavor jets and particle
distributions within jets can be used to constrain parton energy loss mechanisms and
to probe how the ``lost'' energy is redistributed to other partons and the subsequent particles emerging from the collision. In general, these measurements and {analyses} 
 may lead to a better understanding of heavy quark propagation inside the medium~\cite{Wang:2019xey,Nahrgang:2013saa,Cao:2015cba,Hambrock:2017sno}. Such measurements can provide complementary information to the measurements of inclusive heavy-flavor mesons~\cite{Dong:2019byy}, such as the nuclear modification factor~\cite{ALICE:2015vxz,ALICE:2015ccw,CMS:2017qjw,CMS:2017uoy,CMS:2016mah} and azimuthal anisotropy\mbox{~\cite{CMS:2016mah,ALICE:2014qvj,CMS:2017vhp,STAR:2017kkh,ALICE:2017pbx}.} Experiments at RHIC and at the LHC have been performed to investigate the angular correlations of particles associated with heavy-flavor jets. In this section, a brief overview of these measurements is presented. 

\begin{figure}[H]
\includegraphics[scale=0.3]{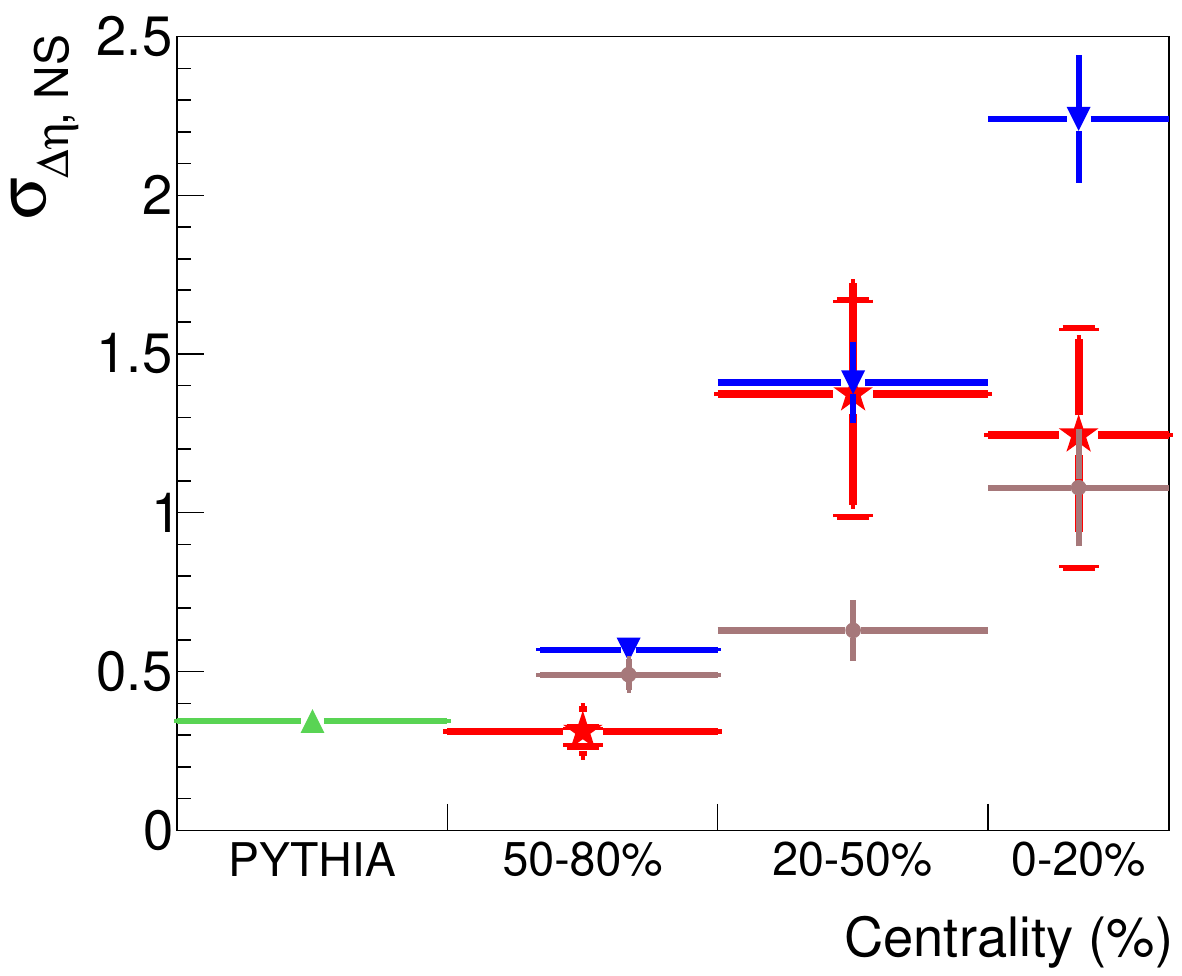}
\includegraphics[scale=0.3]{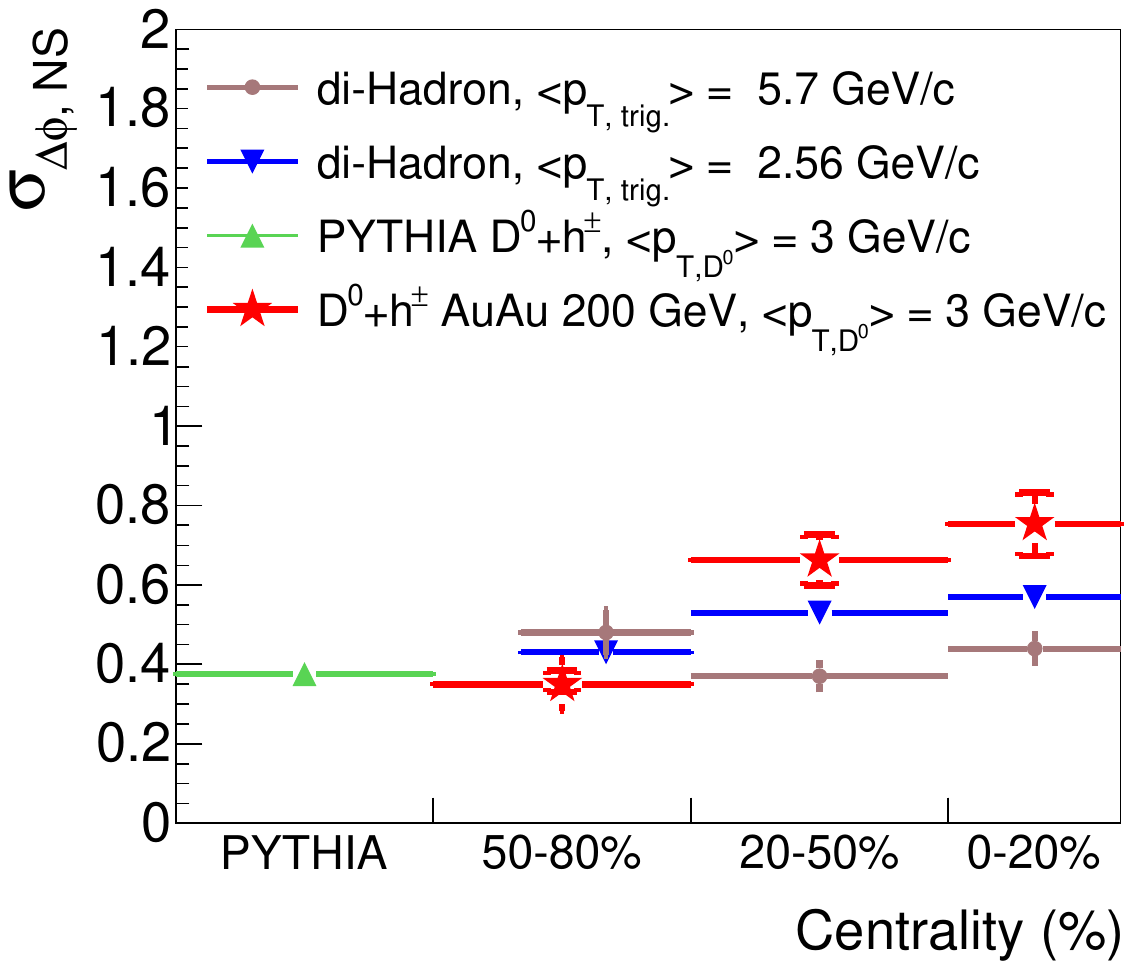}

\centering
 \includegraphics[scale=0.3]{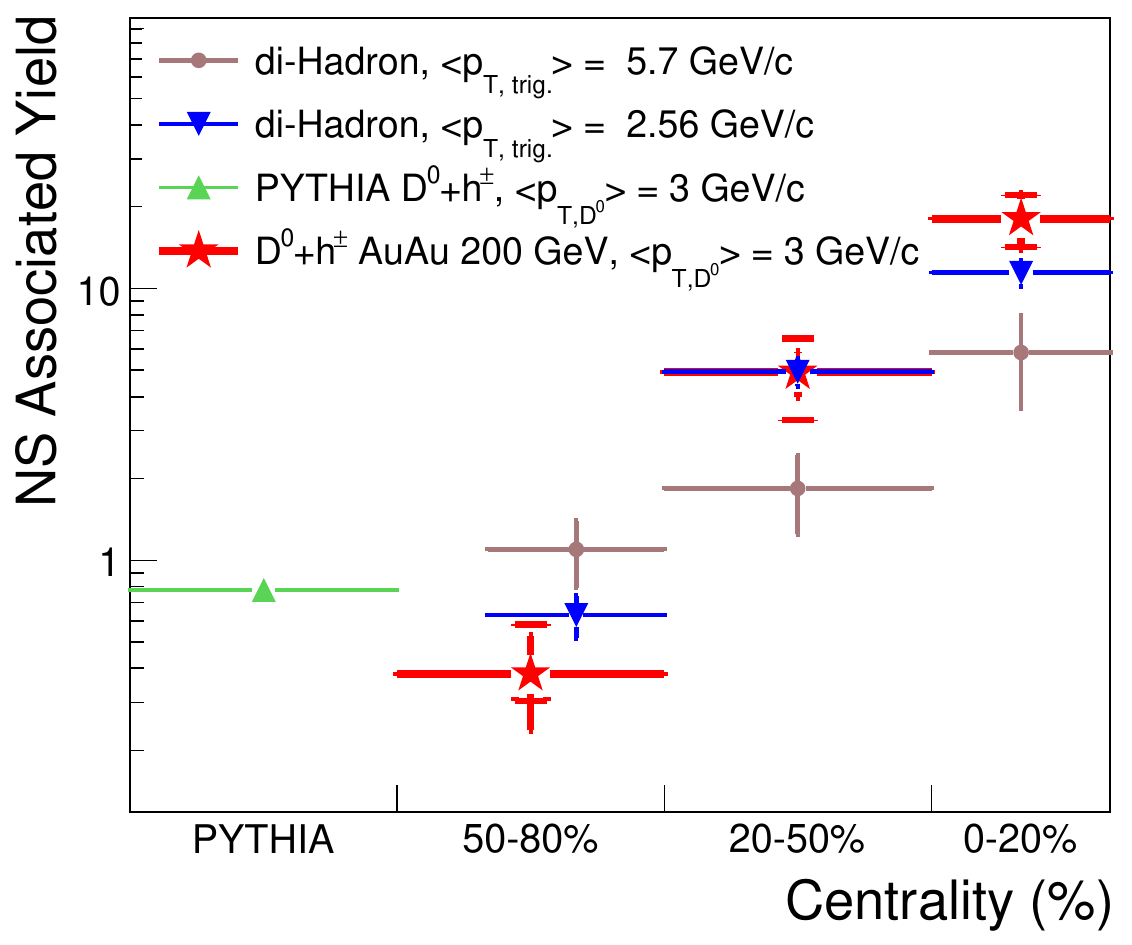}
\caption{Near-side jet-like peak properties of $\rm{D}^{0}$ meson and hadron correlation distribution in Au--Au collisions at $\sqrt{s_{\rm{NN}}} = 200$ GeV, measured by the STAR Collaboration. The near-side peak width along $\Delta\eta$ (\textbf{top left}), width along $\Delta\varphi$ (\textbf{top right}), and correlated hadron yield per $\rm{D}^{0}$ trigger (\textbf{bottom}) are shown. PYTHIA predictions and di-hadron results~\cite{STAR:2011ryj} are also included. Vertical bars show the statistical errors, and cross-bars show the systematic uncertainties~\cite{STAR:2019qbf}.}
\label{fig:Star_DHDPhi-AuAu}
\end{figure}

\begin{figure}[H]

\includegraphics[scale=0.48]{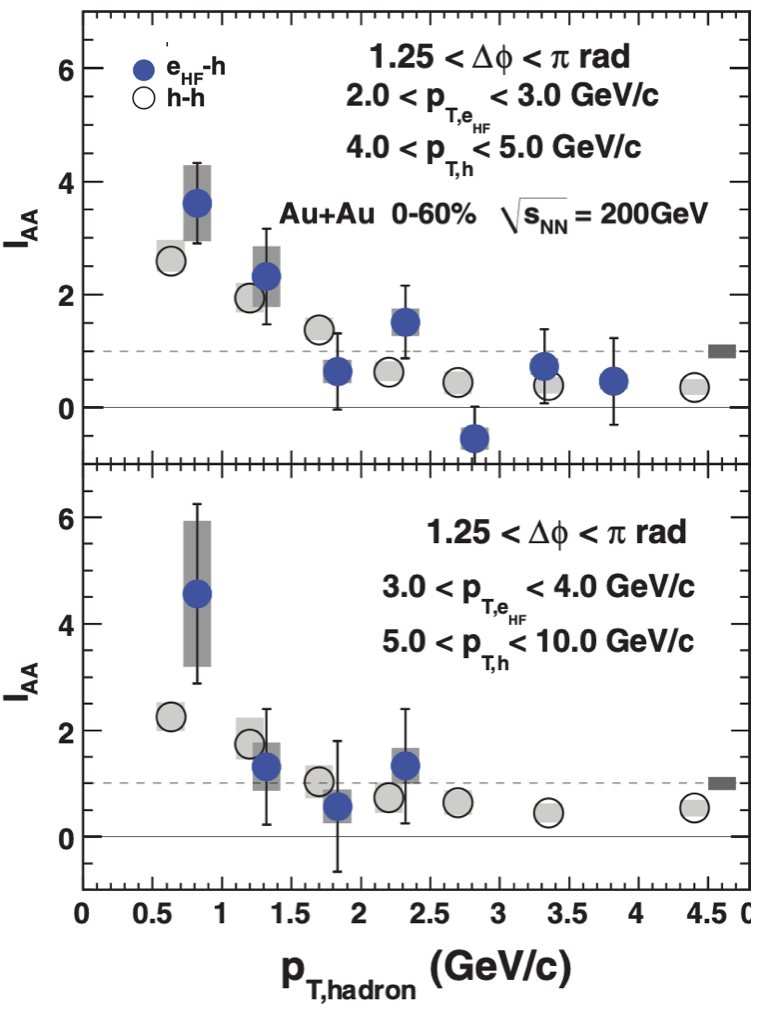}
\caption{$I_{\rm AA}$ determined from the per-trigger yield of away-side $\Delta\varphi$ distribution of electrons from heavy-flavor hadron decays and charged particles in Au--Au collisions at $\sqrt{s_{\rm NN}} = 200$ GeV, measured by the PHENIX Collaboration. The $\Delta\varphi$ range used is $1.25 < \Delta\varphi < \pi$ rad. For comparison, the di-hadron $I_{\rm AA}$ values~\cite{PHENIX:2008osq} are also shown for trigger $p_{\rm T}$ selections where the parent heavy meson has similar $p_{\rm T}$ as the trigger light hadron~\cite{PHENIX:2010cfl}.}
\label{fig:Phenix_HfeHDPhi-AuAu}
\end{figure}

The PHENIX and STAR Collaborations at RHIC performed studies of angular correlations of electrons from heavy-flavor hadron decays with charged hadrons~\cite{PHENIX:2010cfl}, and angular correlations of D mesons with charged hadrons~\cite{STAR:2019qbf}, respectively. In pp collisions, such a correlation distribution is characterized by a jet peak at small $\Delta\varphi$ due to particle pairs from the same fragmentating jet, and a jet peak at $\Delta\varphi \sim \pi$ due to particle pairs from the fragmentating partons in back-to-back di-jet. Angular correlation measurements in pp collisions are discussed in much detail in the previous Section ~\ref{sect:jetfragm}. In nucleus--nucleus collisions, these correlations can provide information about the pattern of energy loss for the back-to-back di-jet system as well as interaction between the fast partons and the medium. The STAR Collaboration performed a study of the centrality dependence of 2D angular correlations ($\Delta\eta,\Delta\varphi$) of $\rm{D}^{0}$ mesons ($2 < p_{\rm{T}}^{\rm{D}^0} < 10$ GeV$/c$), produced by charm quark hadronization after it traverses the medium, and associated charged hadrons ($p_{\rm{T}}$ integrated) in Au--Au collisions at $\sqrt{s_{\rm{NN}}} = 200$ GeV~\cite{STAR:2019qbf}. The main focus of this analysis was the near-side correlation distribution, within $\Delta\varphi \leq \pi/2$, measuring the 2D widths of the jet-like peak, and the number of associated charged hadrons associated with the triggered $\rm{D}^{0}$ meson. The near-side yield and peak widths as a function of the collision centrality are shown in Figure~\ref{fig:Star_DHDPhi-AuAu}. The expectations from PYTHIA Monte Carlo simulations~\cite{Sjostrand:1987su,Shi:2015cva}, as a proxy for pp collisions, are also included. The yields and widths in 50--80\% central Au--Au collisions are consistent with the PYTHIA predictions within the measured uncertainties. The near-side yields and the widths are observed to increase towards more central Au--Au collisions, similar to what was measured for unidentified di-hadron correlations~\cite{STAR:2011ryj}. The increase in the near-side yields and widths in most central collisions is observed for the same $p_{\rm{T}}$ range, where a strong suppression in $\rm{D}^{0}$ meson yield is observed~\cite{STAR:2019qbf}, thus bringing complementary information about charm quark propagation in the QGP medium. The measurement could indicate that the energy lost by the charm quark results in the production of new particles accompanying the D meson. The PHENIX Collaboration measured angular correlations of electrons from heavy-flavor hadron decays and charged particles in Au--Au collisions at $\sqrt{s_{\rm{NN}}} = 200$ GeV in 0--60\% centrality, for two trigger electron $p_{\rm{T}}$ intervals, $2 < p_{\rm{T}}^{\rm{e}} < 3$ GeV$/c$ and $3 < p_{\rm{T}}^{\rm{e}} < 4$ GeV$/c$, and for different associated charged-particle $p_{\rm{T}}$ intervals~\cite{PHENIX:2010cfl}. 
To investigate the possible modification of the jet produced by the opposite heavy quark with respect to one producing the trigger electron, the away-side ($1.25 < \Delta\varphi < \pi$ rad) yield was obtained. The ratio of the away-side yields in Au--Au collisions to pp collisions is shown in Figure~\ref{fig:Phenix_HfeHDPhi-AuAu}. The $I_{\rm{AA}}$ is the largest and above unity for low associated-particle $p_{\rm{T}}$, and decreases with increasing associated particle $p_{\rm{T}}$. The $I_{\rm{AA}}$ obtained for correlations of electrons from heavy-flavor hadron decays is compared with the one obtained for unidentified di-hadron correlations with similar average triggered hadron $p_{\rm{T}}$. The $I_{\rm{AA}}$ for heavy-flavor trigger particles is consistent with that of unidentified charged particles~\cite{PHENIX:2008osq}, though within large uncertainties, which could indicate similar {modifications of charged particles inside light-flavor and heavy-flavor jets }
due to interaction with the QGP medium.

\vspace{-12pt}

\begin{figure}[H]

\includegraphics[scale=0.4]{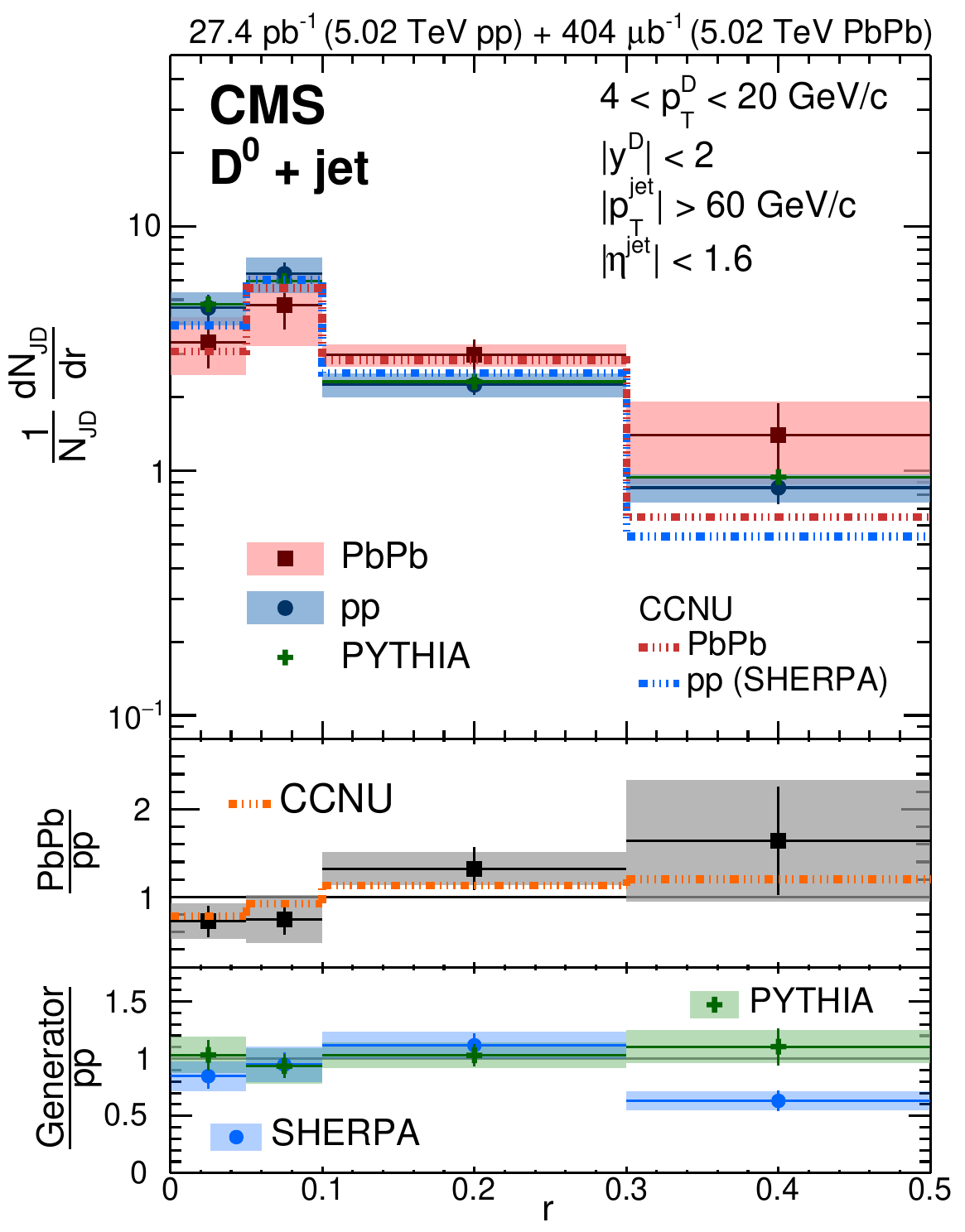}
\caption{Distribution 
 of $\rm{D}^{0}$ mesons ($4 < p_{\rm T}^{D^{0}} < 20$ GeV$/c$) in jets, as a function of the distance from the jet axis ($r$) for $p_{\rm T}^{\rm{jet}} > 60$ GeV$/c$ and $|\eta^{\rm{jet}}|<1.6$ in pp and Pb--Pb collisions at $\sqrt{s_{\rm{NN}}} = 5.02$~TeV from the CMS Collaboration.  The Pb--Pb spectra are compared to the CCNU energy loss model, while the pp spectra are compared with predictions from the PYTHIA and SHERPA pp MC event generators. The ratios of the $\rm{D}^{0}$ meson radial distributions in Pb--Pb to the pp data are shown in the middle panel. In the bottom panel, the ratios of the $\rm{D}^{0}$ meson radial distributions predicted by the two MC event generators to the CMS results in pp collisions are presented~\cite{CMS:2019jis}.}
\label{fig:CMS_D0Jets_PbPb}
\end{figure}

Measurements of angular correlations between heavy-flavor mesons and jets can be used to constrain parton energy loss mechanisms and to better understand the heavy quark diffusion (i.e., propagation) inside the QGP medium~\cite{Wang:2019xey}. The charm quark diffusion with respect to the jet axis was measured by the CMS Collaboration~\cite{CMS:2019jis} in pp and Pb--Pb collisions, for two $\rm{D}^0$ meson $p_{\rm{T}}$ intervals, a lower $p_{\rm{T}}$ interval of $4 < p_{\rm{T}}^{\rm{D}^{0}} < 20$ GeV$/c$ and high $p_{\rm{T}} > 20$ GeV$/c$, for $p_{\rm{T}}^{\rm{Jet}} > 60$ GeV$/c$. The measured observable is the radial distribution of the $\rm{D}^{0}$ mesons with respect to the jet axis, $r = \sqrt{\Delta\varphi^2 + \Delta\eta^2}$, defined as the quadratic sum of the differences in pseudorapidity and azimuth between the $\rm{D}^{0}$ meson and the jet axis direction, shown in Figure~\ref{fig:CMS_D0Jets_PbPb} for $4 < p_{\rm{T}}^{\rm{D}^{0}} < 20$ GeV$/c$. The average value of $r$ for low-$p_{\rm{T}}$ $\rm{D}^0$ mesons was measured to be $0.198 \pm 0.015~(\rm{stat}) \pm 0.005~(\rm{sys})$ and $0.160 \pm 0.007~(\rm{stat}) \pm 0.009~(\rm{sys})$ for Pb--Pb and pp collisions, respectively. This indicates that $\rm{D}^0$ mesons at low $p_{\rm{T}}$ are farther away from the jet axis in Pb--Pb compared to pp collisions. At higher $\rm{p}_{T}$, the radial distribution of $\rm{D}^0$ mesons is more similar in Pb--Pb and pp collisions. The pp results are compared with PYTHIA~\cite{Sjostrand:2014zea} and SHERPA~\cite{Gleisberg:2008ta} event generators, which capture the data trend well within the measured uncertainties. The Pb--Pb distribution is compared to an energy loss model, CCNU~\cite{Wang:2019xey}, which includes in-medium elastic (collisional) and inelastic (radiative) interactions for both the heavy and the light quarks. The model predicts a small depletion of the $\rm{D}^0$ meson yield at small $r$ and an enhancement of yield at larger $r$ compared to pp collisions, which is consistent with the trend seen in the data, as can be seen in the ratio plot of Figure~\ref{fig:CMS_D0Jets_PbPb}. This measurement hints at a modification of the $\rm{D}^0$ meson radial profile in Pb--Pb collisions at low $p_{\rm{T}}$, possibly induced by interactions of the charm quark with the medium constituents, which alter the original quark direction. This modification vanishes at higher $p_{\rm{T}}$. 
\vspace{-4pt}

\begin{figure}[H]

\includegraphics[scale=0.68]{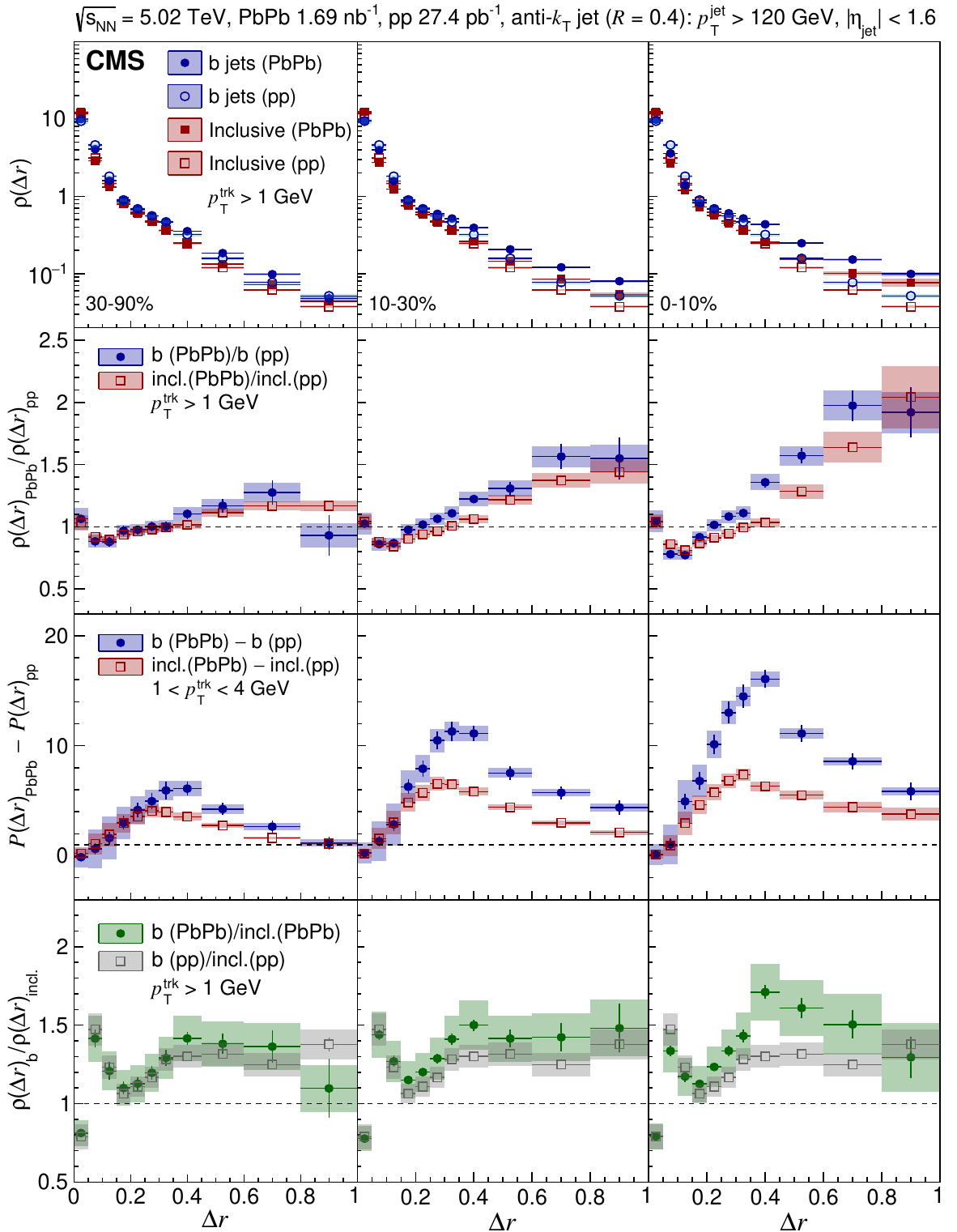}
\caption{\textbf{First row
}: jet shape distributions $\rho(\Delta r)$ for inclusive and b-jets with $p_{\rm{T}} > 120$ GeV$/c$ in three centrality bins (left to right) for Pb--Pb collisions at $\sqrt{s_{\rm{NN}}} = 5.02$ TeV, and in pp collisions, both measured by the CMS Collaboration. \textbf{Second row}: ratio of Pb--Pb to pp jet shape results for inclusive jets (red) and b-jets (blue). \textbf{Third row}: difference between the charged-particle transverse momentum profile between Pb--Pb and pp collisions for inclusive and b-jets. \textbf{Fourth row}: ratio of b- to inclusive jet shapes for several Pb--Pb centrality bins (green), as well as pp collisions (identical in all three panels). In all panels, the vertical bars and shaded boxes represent the statistical and systematic uncertainties, respectively~\cite{CMS:2022btc}.} 
\label{fig:CMS_BJets_PbPb}
\end{figure}

Similarly, correlation measurements can be used to study the modification of jet shapes in heavy-ion collisions, using charged hadron constituents as a function of their radial distance from the jet axis, as performed by the CMS experiment~\cite{CMS:2022btc}. These measurements can give insight into details of jet quenching and medium response to the evolving jet. The CMS experiment used jets initiated by beauty quarks (b-jets) to provide unique experimental means to investigate the mass dependence of quenching effects and parton shower evolution. The transverse momentum profile $P(\Delta r)$ of charged particles in the jets, defined as $P(\Delta r) = \frac{1}{\Delta r_{\rm{b}} - \Delta r_{\rm{a}}} \frac{1}{N_{\rm{jet}}} \sum_{\rm{jets}}$$\sum_{ \rm{trk}}$$_{\in(\Delta r_{\rm{a}}, \Delta r_{\rm{b}})} p_{\rm{T}}^{\rm{trk}}$, 
was measured, where $\Delta r = \sqrt{\Delta\varphi^2 + \Delta\eta^2}$ is the radial distance between a track and the jet axis defined in pseudorapidity and azimuthal angle, $\Delta r_{\rm{a}}$ and $\Delta r_{\rm{b}}$ are the edges of rings in $\Delta r$, and $p_{\rm{T}}^{\rm{trk}}$ is the charged particle's transverse momentum. The $P(\Delta r)$ distribution is normalized to unity within $\Delta r < 1$ to produce the jet shape distribution, $\rho(\Delta r)$, that indicates how the momentum of charged particles is distributed with respect to the jet axis. The shapes of b-jets and inclusive jets of $p_{\rm{T}} > 120$~GeV$/c$ for charged particles with $p_{\rm{T}} > 1$ GeV$/c$ was measured for Pb--Pb and pp collisions, as shown in Figure~\ref{fig:CMS_BJets_PbPb}. The b-jets are found to be broader than inclusive jets. The ratio of $\rho(\Delta r)$ distribution in Pb--Pb to pp collisions ($2^{\rm{nd}}$ row panels) shows a depletion of particles for low $\Delta r$ and a strong enhancement at large $\Delta r$, indicating redistribution of $p_{\rm{T}}$ of jet constituents from small to large distances from the jet axis. The large $\Delta r$ enhancement in Pb--Pb collisions is centrality-dependent and is most significant in central collisions, indicating a modification of energy flow around the jet axis in the presence of the QGP medium. The large $\Delta r$ enhancement in Pb--Pb collisions is more pronounced for b-jets than inclusive jets, showing mass dependent interactions in the QGP. The difference in the transverse momentum profile between Pb--Pb and pp collisions characterizes the magnitude of the measured excess momentum as shown in the third row of the figure. A more significant transverse momentum excess in Pb--Pb collisions at intermediate and high $\Delta r$ is found for b-jets than for inclusive jets. A comparison of b-jet shapes to those of inclusive jets is shown in the bottom panel for pp and Pb--Pb collisions. In pp collisions, b-jets show a depletion for $\Delta r < 0.05$, with respect to inclusive jets, that could be interpreted as being due to the dead-cone effect~\cite{Dokshitzer:1991fd} (suppression of collinear parton radiation from a massive emitter such as a heavy quark). In Pb--Pb collisions, the depletion at small $\Delta r$ is similar to pp collisions. For higher $\Delta r$, b-jet shapes are broader than inclusive jet shapes in pp and Pb--Pb collisions, with a significant enhancement in the most central Pb--Pb collisions. This measurement provides new constraints for theoretical calculations of parton flavor dependence of energy loss and jet--medium interactions in the quark--gluon plasma.

\begin{figure}[H]

\includegraphics[scale=1.1]{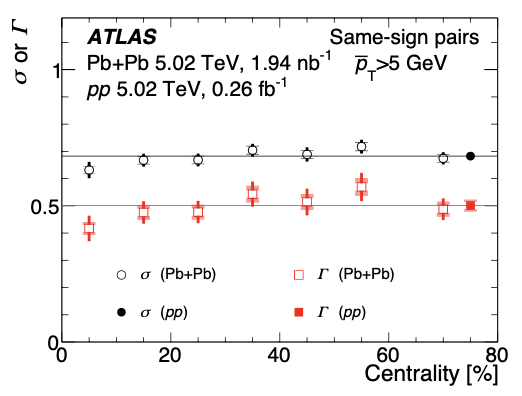}
\caption{The 
 measured widths of the away-side peak of the angular correlations between same-sign muon pairs in pp and Pb--Pb collisions at $\sqrt{s_{\rm{NN}}} = 5.02$ TeV from the ATLAS Collaboration.  The vertical bars and shaded bands on the data points represent the statistical uncertainties, and systematic uncertainties, respectively. The horizontal lines indicate the nominal pp values plotted across the full centrality range~\cite{ATLAS:2023vms}.} 
\label{fig:ATLAS_MuMuDphiWidth_PbPb}
\end{figure}

At momenta comparable to or smaller than the quark mass, heavy quarks are thought to undergo Brownian-like motion in the quark--gluon plasma, with their transport characterized by a diffusion coefficient~\cite{Casalderrey-Solana:2006fio}. 
The process of losing energy via gluon radiation is referred to as radiative energy loss~\cite{Djordjevic:2003zk}. Heavy quarks, when produced at LO, will have a back-to-back correlation in azimuthal angle between the quark and anti-quark, due to momentum conservation. As a consequence of the multiple interactions with the medium, this initial correlation can broaden around $\Delta\varphi = \pi$. Energy loss via a radiative mechanism may dampen this broadening~\cite{Zakharov:2020sfx}. 
Angular correlations of heavy quarks can thus be sensitive to the relative contribution of collisional and radiative energy loss processes~\cite{Nahrgang:2013saa,Cao:2015cba}. 
The ATLAS Collaboration performed measurements of angular correlations between muon pairs in pp and Pb--Pb collisions at $\sqrt{s_{\rm{NN}}} = 5.02$~TeV~\cite{ATLAS:2023vms}. Muon pairs from beauty hadron decays were selected by considering muons with the same charge with both muons having $p_{\rm{T}} > 4$~GeV$/c$. The $\Delta\varphi$ distribution of muon pairs shows a clear peak on the away-side, consistent with the back-to-back configuration of beauty quark pair production from hard scattering processes. The width of the away-side peaks was characterized by fitting the $\Delta\varphi$ distribution with a Cauchy--Lorentz function. The centrality dependence of the away-side width was measured, as shown in Figure~\ref{fig:ATLAS_MuMuDphiWidth_PbPb}. The widths obtained for different centralities in Pb--Pb collisions are consistent with the value measured in pp collisions, with a slightly reduced value observed for the 0--10\% centrality interval. To further investigate the mechanisms by which a heavy quark interacts with the QGP constituents, this measurement can be extended to different $p_{\rm{T}}$ regions of muon pairs. Comparison with model calculations including collisional and/or radiative energy loss would help with the interpretation of the current observation. The measurement will provide important constraints on theoretical descriptions of the dynamics of heavy quarks inside the quark--gluon plasma.

\vspace{-4pt}

\section{Small-System Collective-Like Effects for Heavy Quarks}
\label{sect:collectivity}

As discussed in the previous section, Section~\ref{sect:corrHIC}, the properties of the QGP are studied by analyzing high-energy heavy-ion collisions~\cite{PHENIX:2004vcz, STAR:2005gfr,PHOBOS:2004zne,BRAHMS:2004adc,ALICE:2010yje,ALICE:2022wpn}. One of the key signatures of the formation of the QGP in these collision systems is the azimuthal anisotropy of the produced particles~\cite{Voloshin:2008dg}, due to the onset of collective motion of the system that is driven by the specific geometrical structure of the overlap region of the two colliding nuclei. During the medium expansion, the initial-state spatial anisotropy is translated into a momentum anisotropy of the particles emerging from the medium~\cite{Qin:2010pf}. The magnitude of the azimuthal anisotropies is quantified via a Fourier decomposition of the particle azimuthal distribution~\cite{Voloshin:1994mz}, where the Fourier coefficients $v_n$ characterize the strength of the anisotropy. For non-central A--A collisions, where the overlap region typically has an almond shape, the largest contribution to the azimuthal anisotropy is provided by the second-order Fourier coefficient $v_2$, referred to as the elliptic flow coefficient. Its value is used to characterize the strength of the collective motion of the system. In two-particle angular correlation distributions measured in non-central A--A collisions, the effect of the elliptic flow can be seen as pronounced structures on the near and away sides along $\Delta\varphi$, extending over a large $\Delta\eta$ region, which are commonly referred to as ``ridges''~\cite{Abelev:2009af}. The measurements are well described by models invoking a hydrodynamic expansion of the hot and dense asymmetrical medium produced in the collision. Surprisingly, similar long-range ridge structures and a positive $v_2$ coefficient were also observed for light-flavor particles in high-multiplicity pp and p--Pb collisions at the LHC~\cite{Abelev:2012ola,Aaboud:2016yar,Chatrchyan:2013nka,ABELEV:2013wsa,Khachatryan:2014jra,Khachatryan:2010gv}, and in high-multiplicity $d-$Au and  ${}^{3}$He--Au collisions at RHIC~\cite{Adare:2013piz,Adamczyk:2015xjc}. The interpretation of the positive $v_2$ in these small systems is currently highly debated~\cite{Loizides:2016tew}. It has raised the question of whether a fluid-like QGP medium with a size smaller than that produced in A--A collisions is created~\cite{Werner:2010ss, Deng:2011at}. Alternate explanations foresee mechanisms involving initial-state effects, such as gluon saturation within the color glass condensate effective field theory~\cite{Dusling:2013qoz, Bzdak:2013zma}, or final-state color--charge exchanges~\cite{Dumitru:2013tja,Wong:2011qr}. 

Heavy quarks, produced during the early stages of hadronic collisions~\cite{Cacciari:1998it, Cacciari:2001td, Kniehl:2005ej, Kniehl:2007erq}, can be used to probe both initial- and final-state effects of the collision dynamics~\cite{Bierlich:2018xfw,Dusling:2015gta,Zhang:2019dth,Zhang:2020ayy,Dong:2019byy}. In A--A collisions, strong elliptic flow signals were observed for leptons from the decay of heavy-flavor hadrons and open-charm D mesons~\cite{Adam:2015pga,Adam:2016ssk,Acharya:2017qps, ALICE:2014qvj, CMS:2017vhp,ALICE:2022wpn}, suggesting that charm quarks develop significant collective behavior via their strong interactions with the bulk of the QGP medium. Measurements of elliptic flow of hidden-charm J$/\psi$ mesons provide further evidence for strong rescatterings of charm quarks~\cite{ALICE:2017quq}. Recent measurements of positive $v_2$ for non-prompt D mesons (i.e., D mesons produced from the decays of beauty hadrons), though with smaller values than prompt D meson $v_2$, were released by the CMS~\cite{CMS:2022vfn} and ALICE Collaborations~\cite{ALICE:2023gjj}, suggesting that beauty quarks could also participate in the medium collective motion, though with lesser extent than charm quarks.

In small colliding systems, the study of heavy-flavor hadron collectivity has the potential to disentangle \textls[-15]{possible contributions from both initial- and final-state effects~\mbox{\cite{Zhang:2019dth,Li:2018leh,Zhang:2020ayy}.}} In particular, heavy-flavor hadrons may be more sensitive to possible initial-state gluon saturation effects. The Collaboration at the LHC performed several measurements to evaluate the $v_2$ of charm and beauty hadrons in pp and p--Pb collisions, using two-particle angular correlation techniques~\cite{ALICE:2017smo, ALICE:2018gyx, CMS:2018duw, ATLAS:2019xqc, CMS:2020qul, ALICE:2022ruh}. The general procedure performed was to obtain the angular correlations of heavy-flavor trigger particles with charged particles~\cite{ALICE:2018gyx} in events with high multiplicity. The correlation distribution in these events contains contributions related to collectivity and jet fragmentation, with the latter being referred to as non-flow effects. These non-flow contributions can be suppressed by requiring a pseudorapidity separation ($\Delta\eta$) between heavy-flavor particles and charged particles. The azimuthal correlation distribution normalized to the number of trigger particles is obtained. While a large $\Delta\eta$ separation largely reduces non-flow contributions, especially for the near-side region, a significant contribution from recoil jet fragmentation still remains on the away-side of the $\Delta\varphi$ distribution. This can be subtracted by using the azimuthal correlation distributions measured in low-multiplicity events. The subtraction method relies on the assumptions that the jet correlations on the away side remain unmodified as a function of the event multiplicity and that there are no significant correlations due to collective effects in low-multiplicity collisions. The non-flow subtracted $\Delta\varphi$ distribution is fit with a Fourier decomposition, $a [1 + 2V_{1\Delta}^{\rm{HF-ch}} \cos(\Delta\varphi) + 2 V_{2\Delta}^{\rm{HF-ch}} \cos(2\Delta\varphi)]$. The second-order coefficient $2V_{2\Delta}^{\rm{HF-ch}}$, which is the dominant term, is obtained. Using the assumption that $V_{2\Delta}^{\rm{HF-ch}}$ can be factorized as a product of single-particle $v_2$ coefficients, the elliptic flow coefficient of the heavy-flavor particle, $v_2^{\rm{HF}}$, is extracted from the equation $v_2^{\rm{HF}} = V_{2\Delta}^{\rm{HF-ch}} / v_2^{\rm{ch-ch}}$. 
The $v_2$ values of several heavy-flavor particle species in small systems were measured by the ALICE Collaboration using leptons from heavy-flavor hadron decays~\cite{ALICE:2018gyx, ALICE:2022ruh} and J$/\psi$~\cite{ALICE:2017smo}, by the ATLAS Collaboration using muons from charm and beauty hadron decays~\cite{ATLAS:2019xqc}, and by the CMS Collaboration using prompt $\rm{D}^0$~\cite{CMS:2018loe}, non-prompt $\rm{D}^0$ mesons~\cite{CMS:2020qul}, and J$/\psi$~\cite{CMS:2018duw} mesons. 

\begin{figure}[H]
\includegraphics[scale=0.33]{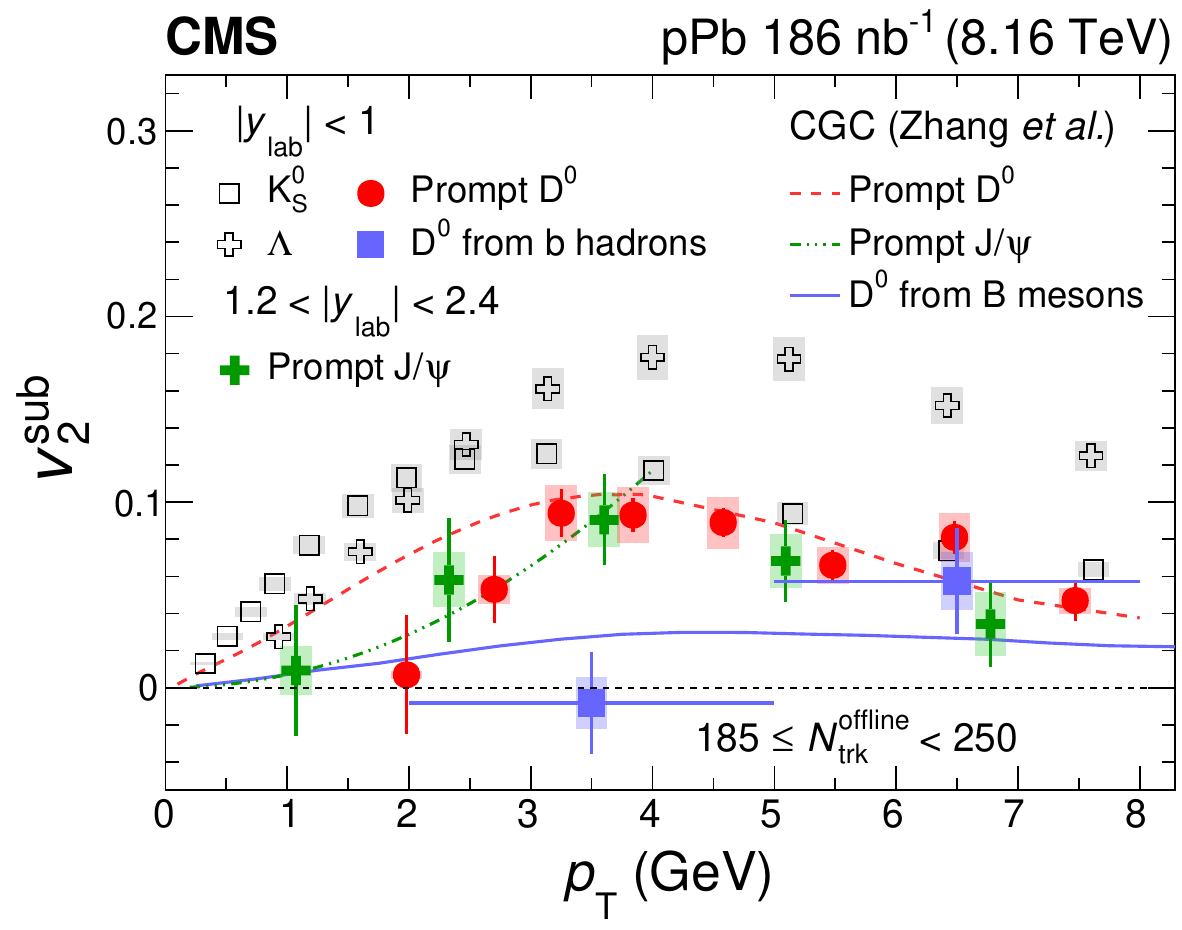}
\includegraphics[scale=0.36]{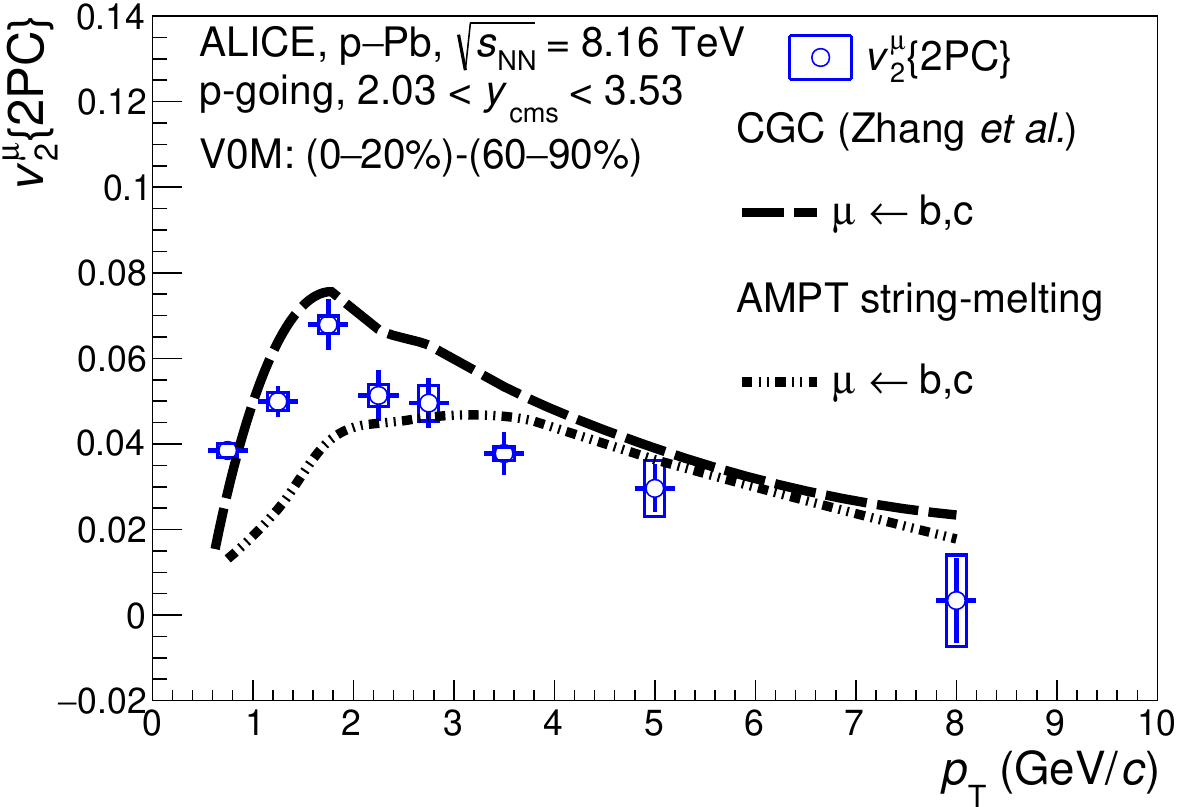}
\caption{(\textbf{Left
}) $v_2$ of prompt~\cite{CMS:2018loe} and non-prompt $\rm{D}^0$ mesons~\cite{CMS:2020qul}, $\rm{K}^0_s$ mesons and $\Lambda$ baryons at mid-rapidity~\cite{CMS:2018loe}, and prompt J$/\psi$ mesons at forward rapidity~\cite{CMS:2018duw}, measured by the CMS Collaboration as a function of $p_{\rm{T}}$ in high-multiplicity p--Pb collisions at $\sqrt{s_{\rm{NN}}}=8.16$ TeV. (\textbf{Right}) The $p_{\rm{T}}$ differential $v_2^{\mu}$ of inclusive muons measured by the ALICE Collaboration at forward rapidity in high-multiplicity p--Pb collisions at $\sqrt{s_{\rm{NN}}}=8.16$ TeV~\cite{ALICE:2022ruh}, compared with predictions of $v_2$ of muons from heavy-flavor hadron decays from CGC and AMPT models. The vertical bars correspond to the statistical uncertainties, while the boxes denote the systematic uncertainties.} 
\label{fig:HFv2_pPb}
\end{figure}

The $v_2$ of prompt $\rm{D}^0$ mesons~\cite{CMS:2018loe}, non-prompt $\rm{D}^0$ mesons~\cite{CMS:2020qul}, and prompt J$/\psi$~\cite{CMS:2018duw} as a function of $p_{\rm{T}}$ for p--Pb collisions at $\sqrt{s_{\rm{NN}}}=8.16$ TeV, measured by the CMS Collaboration, is shown in Figure~\ref{fig:HFv2_pPb} (left panel). 
The $v_2$ of strange hadrons~\cite{CMS:2018loe} is also shown. A positive $v_2$ value for prompt $\rm{D}^0$ and J$/\psi$ is observed, with a rising and then a decreasing trend with $p_{\rm{T}}$, peaking at about 3--4 GeV/$c$. A clear ordering in the $v_2$ values is observed for the low-$p_{\rm{T}}$ region ($p_{\rm{T}} < 3$ GeV$/c$), where heavier particles have smaller $v_2$ at a given $p_{\rm{T}}$ value. A similar mass ordering for the $v_2$ of $\rm{D}^0$ mesons and strange hadrons in semi-central Pb--Pb collisions is observed, although the multiplicity range in Pb--Pb collisions is much larger~\cite{CMS:2017vhp}. In Pb--Pb collisions, this behavior is understood to be due to particle emission from a collectively expanding source with a common velocity field. This might indicate a significant collective behavior of charm quarks in high-multiplicity p--Pb systems at LHC energies. For non-prompt 
$\rm{D}^0$ mesons, the $v_2$ values are consistent with zero at low $p_{\rm{T}}$, while at high $p_{\rm{T}}$, a hint of a positive $v_2$ value is observed, but it is not significant within statistical and systematic uncertainties. The non-prompt $\rm{D}^0$ $v_2$ is observed to be smaller than that of prompt $\rm{D}^0$ mesons with a significance of 2.7 standard deviations. This also indicates a mass hierarchy of the original quark participating in the collective-like dynamics. In Pb--Pb collisions, a similar mass ordering for muon $v_2$ from charm and beauty decay at low $p_{\rm{T}}$ was observed~\cite{ATLAS:2020yxw}, and is understood as being due to final-state scattering mechanisms~\cite{Nahrgang:2013xaa,Ke:2018jem,Katz:2019fkc}. The values of $v_2$ from a calculation within the color glass condensate (CGC) framework~\cite{Dusling:2015gta,Zhang:2019dth,Zhang:2020ayy} for prompt J$/\psi$ mesons, prompt and non-prompt $\rm{D}^0$ mesons in p--Pb collisions are compared with the data in Figure~\ref{fig:HFv2_pPb} (left panel). Within the CGC framework, correlations in the initial stage of the collision between partons originating from projectile protons and dense gluons in the lead nucleus can generate a sizable elliptic flow. The model qualitatively describes the data, suggesting that initial-state effects may play an important role in the generation of collectivity for these particles in p--Pb collisions. The CGC framework also predicts a mass hierarchy between prompt and non-prompt $\rm{D}^0$ mesons for $p_{\rm{T}} \sim$ 2--5 GeV$/c$, consistent with the data within uncertainties.

The ALICE Collaboration measured the $v_2$ of muons at forward ($2.03 < y_{\rm{cms}} < 3.53$) and backward rapidity ($-4.46 < y_{\rm{cms}} < -2.96$) in high-multiplicity p--Pb collisions at $\sqrt{s_{\rm{NN}}} = 8.16$ TeV~\cite{ALICE:2022ruh}. For $p_{\rm{T}} > 2$ GeV$/c$, a dominant contribution of muons is produced from heavy-flavor hadron decays. A positive $v_2$ was measured at both rapidities. To better understand the source of the observed azimuthal anisotropies in small collision systems, the measurement was compared with a multi-phase transport (AMPT) model~\cite{Lin:2004en,Li:2018leh,Lin:2021mdn} and CGC~\cite{Dusling:2015gta,Zhang:2019dth,Zhang:2020ayy} model calculations for muons from heavy-flavor hadron decays. The results for muons measured at forward rapidity are shown in the right panel of Figure~\ref{fig:HFv2_pPb}. The AMPT model provides a microscopic evolution of parton interactions, including a parton escape mechanism described via a parton cascade model~\cite{Zhang:1997ej}. The AMPT model generates a positive $v_2$, mainly driven by the anisotropic parton escape mechanism, where partons have a higher probability to escape the interaction region along its shorter axis~\cite{He:2015hfa}. In the CGC calculations, the correlations in the initial stage of the collision between partons in the colliding proton and gluons in the dense Pb ion generate a significant $v_2$ signal that persists till the final state and is observed in the heavy-flavor hadron decay muon measurement. The CGC-based calculations provide a larger $v_2$ compared to AMPT calculations at low $p_{\rm{T}}$, up to 3 GeV$/c$, while the two models provide similar results and describe the data at high $p_{\rm{T}}$, where heavy-flavor hadron decays dominate the muon sample. This comparison indicates that both initial- and final-state effects can explain the azimuthal anisotropies observed in small collision systems.

The $v_2$ of J$/\psi$ at forward and backward rapidity in high-multiplicity p--Pb collisions was compared with measurements in non-central Pb--Pb collisions by the ALICE experiment~\cite{ALICE:2017smo}. The $p_{\rm{T}}$-dependent $v_2$ values in p--Pb collisions are consistent with those measured in Pb--Pb collision within uncertainties. In Pb--Pb collisions, at low $p_{\rm{T}}$, the $v_2$ coefficient is believed to originate from the recombination of charm quarks thermalized in the medium, which is described by thermal models~\cite{Du:2015wha}. In p--Pb collisions, the amount of produced charm quarks is small and, therefore, the contribution from recombination should be negligible. For $p_{\rm{T}} > 4$ GeV$/c$, the $v_2$ in Pb--Pb collisions is expected to come from path length-dependent suppression inside the QGP medium. In p--Pb collisions, the medium, if any, is expected to have a much smaller size, and hence, very feeble path length-dependent effects are expected. 

\begin{figure}[H]
\includegraphics[scale=0.33]{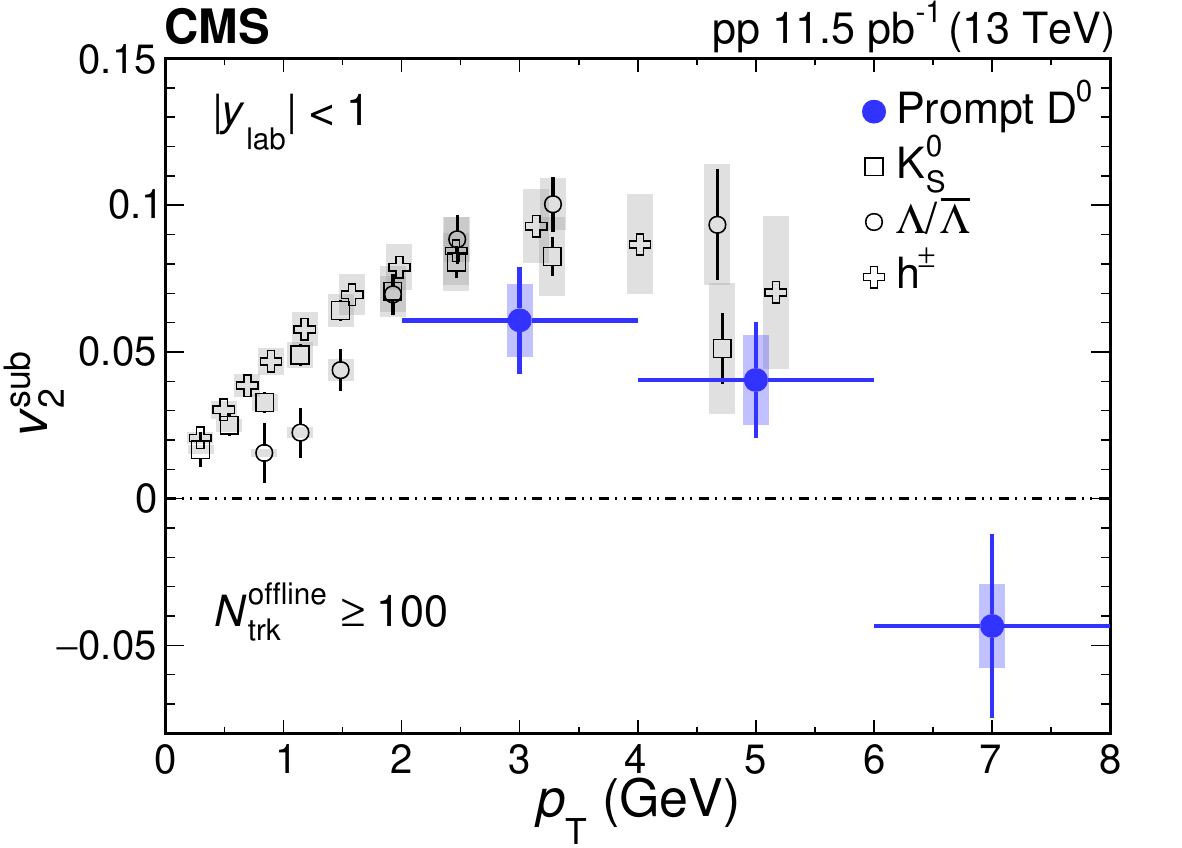}
\includegraphics[scale=0.34]{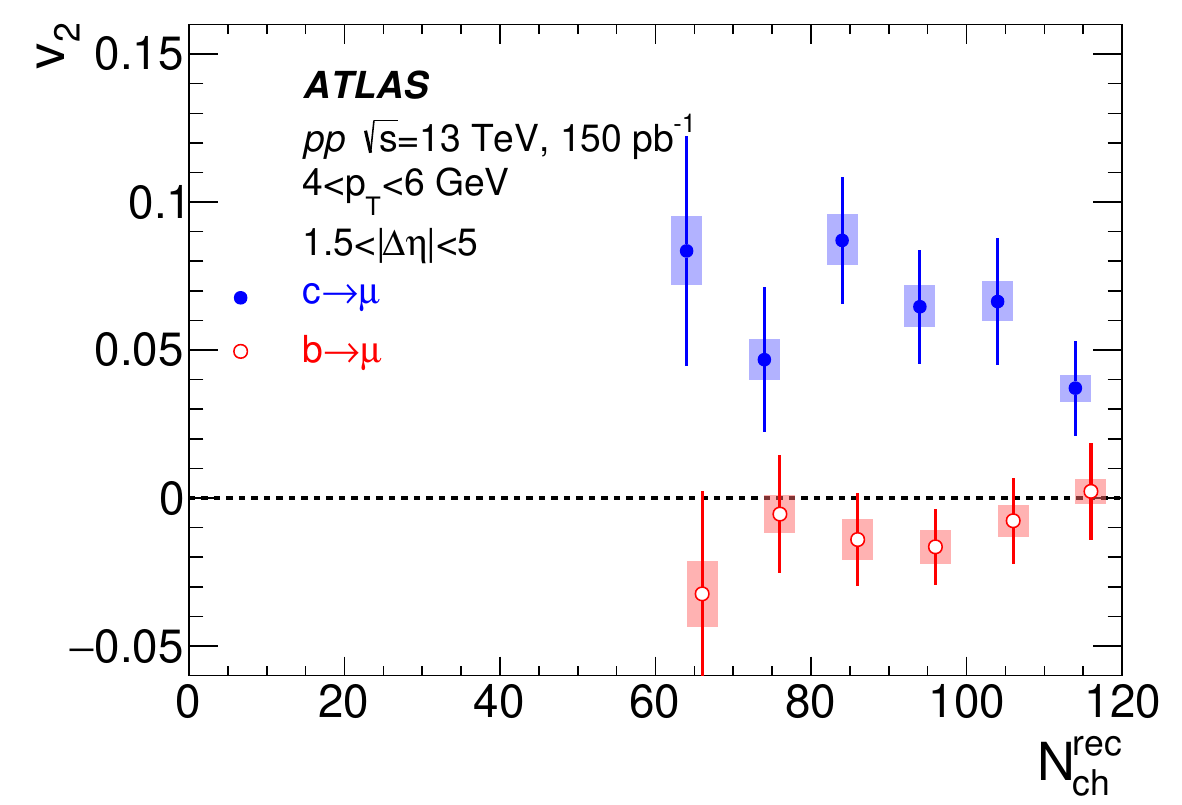}
\caption{(\textbf{Left
}) $v_2$ of prompt $\rm{D}^0$ mesons measured as a function of $p_{\rm{T}}$ at mid-rapidity in high-multiplicity pp collisions at $\sqrt{s}=13$ TeV~\cite{CMS:2020qul}, measured by the CMS Collaboration. The prompt $\rm{D}^0$ meson measurement is compared with those of charged particles, $\rm{K}^0_s$ mesons and $\Lambda$ baryons~\cite{CMS:2016fnw}. (\textbf{Right}) $v_2$ of muons from charm and beauty hadron decays as a function of track multiplicity for muons with $4 < p_{\rm{T}} < 6$ GeV$/c$ in pp collisions at $\sqrt{s}=13$ TeV~\cite{ATLAS:2019xqc}, measured by the ATLAS Collaboration. The vertical bars correspond to the statistical uncertainties, while the shaded areas denote the systematic uncertainties.} 
\label{fig:HFv2_pp}
\end{figure}

The $v_2$ of prompt $\rm{D}^0$ mesons as a function of $p_{\rm{T}}$ in pp collisions at $\sqrt{s}=13$ TeV, measured by the CMS Collaboration~\cite{CMS:2020qul}, is shown in Figure~\ref{fig:HFv2_pp} (left panel). A positive $v_2$ signal for prompt charm hadrons over a $p_{\rm{T}}$ range up to 6 GeV$/c$ is observed, with a decreasing trend towards higher $p_{\rm{T}}$ values. The $v_2$ for prompt $\rm{D}^0$ mesons is found to be comparable, within uncertainties, with that of light-flavor hadron species, i.e., unidentified charged particles (dominated by pions), $\rm{K}_s^0$ mesons and $\Lambda$ baryons~\cite{CMS:2016fnw}, that are also presented in the same figure. At similar event multiplicities, the prompt $\rm{D}^0$ meson $v_2$ values are found to be comparable within uncertainties in pp and p--Pb systems.

The elliptic flow of muons from the decay of charm and beauty hadrons in pp collisions at $\sqrt{s}=13$ TeV was measured by the ATLAS Collaboration~\cite{ATLAS:2019xqc} for $4 < p_{\rm{T}} < 7$~GeV$/c$ and $|\eta|<2.4$. It is shown as a function of reconstructed track multiplicity in Figure~\ref{fig:HFv2_pp} (right panel). A significant non-zero $v_2$ is observed for muons from charm hadron decays, without significant dependence on multiplicity. The $v_2$ of muons from charm hadron decays decreases with increasing $p_{\rm{T}}$, and is consistent with zero for $p_{\rm{T}} \gtrsim 6$ GeV/$c$. The $v_2$ for muons from beauty hadron decays was measured to be consistent with zero within uncertainties through all the multiplicity ranges measured, and also shows no $p_{\rm{T}}$ dependence. These results also indicate a mass hierarchy of the $v_2$ signal at the partonic~level.

The presented measurements of heavy-flavor $v_2$ in high-multiplicity pp and p--Pb collisions, the indication of mass dependence of $v_2$ in p--Pb collisions, and the comparison with models capable of describing the measurements, can provide insights into the origin of heavy-flavor quark collectivity in small colliding systems. However, the identification of the source of the observed collective-like effects remains a topic of debate.
\section{Summary}
\label{sec:Summary}

This article reviews the most recent experimental measurements of correlated heavy-flavor particle production as a function of the relative azimuthal angle and/or rapidity. Angular correlation techniques are used to study the production, fragmentation, and hadronization of heavy quarks in pp collisions and to understand how these processes are modified in the presence of a strongly interacting quark--gluon plasma in heavy-ion collisions. Azimuthal anisotropy observations in high-multiplicity pp and p--A collisions and their possible origin are also reviewed. 

In pp collisions, angular correlations of heavy-flavor particle pairs allow for testing pQCD calculations at different order in $\alpha_{\rm s}$. Such correlation distributions were measured at RHIC and at the LHC through different experiments, and were compared with predictions from different Monte Carlo generators, such as PYTHIA, POWHEG, HERWIG, MADGRAPH, and SHERPA. It is challenging for these models to provide an optimal description of the data. These kinematic correlation observables provide sensitivity to the underlying differences within the models. The shape of the correlation distribution is also used to study the multiple production of heavy quarks in a single pp and p--A collision. The angular separation between heavy-flavor particle pairs shows distinct structural differences when produced in single or double parton scattering processes. 

A thorough characterization of the in-vacuum heavy quark fragmentation process can be performed by measuring the angular correlation distribution between a trigger heavy-flavor particle and associated charged particles, as performed by the ALICE Collaboration. The angular correlation distribution and more quantitative observables extracted from it, such as the near- and away-side peak yields and widths measured in pp and p--Pb collisions, were compared with different Monte Carlo generators such as PYTHIA, POWHEG+PYTHIA, HERWIG, and EPOS. The first two models provide predictions that are closest to the data, but further configuration and parameter tuning could be helpful for a better reproduction of the properties of the correlation distribution. By comparing the correlation distributions in pp and p--Pb collisions, effects from cold nuclear matter on the heavy quark fragmentation process can be studied. No significant impact was observed in the kinematic ranges measured in the experiments. It is also important that the fragmentation studies can be performed as a function of event multiplicity, as correlation techniques used to evaluate flow coefficients in small systems rely on the assumption that jet fragmentation is independent of event multiplicity. Results of correlation measurements as a function of charged-particle multiplicity in pp and p--Pb collisions indicate that, within the experimental uncertainties, the correlation distributions and their properties are consistent for all multiplicity ranges studied, implying a similar fragmentation of charm quarks into final-state mesons.   

In ultra-relativistic heavy-ion collisions, heavy-flavor jets and particle distributions within jets are excellent tools for characterizing heavy quark propagation and to constrain energy loss mechanisms affecting partons traversing the quark--gluon plasma. The STAR and PHENIX experiments at RHIC examined the angular correlations between a trigger heavy-flavor particle and associated charged particles in Au--Au collisions and compared the results with those obtained in pp collisions and with predictions from the PYTHIA event generator. The trigger-particle $p_{\rm{T}}$-integrated near-side yields and widths were observed to increase towards more central Au–Au collisions. The measurement indicates that charm quarks lose energy in the QGP, and the lost energy is converted into additional low-$p_{\rm{T}}$ particles accompanying the charm meson. On the away side, a higher yield of low-$p_{\rm{T}}$-associated particles was observed in Au--Au collision compared to pp, which decreased and hinted towards a suppression for increasing the associated particle $p_{\rm{T}}$. The charm quark diffusion inside the QGP medium, studied at the LHC by the CMS Collaboration, implies that low-$p_{\rm{T}}$ D mesons in Pb--Pb collisions tend to be further displaced from the jet axis compared to pp collisions, due to interaction with the medium constituents. Studies of the modification of jet shapes using correlation techniques, by measuring the distribution of charged particles inside a jet as a function of their radial distance from the jet axis, was performed by the CMS Collaboration for jets initiated by beauty quarks (b-jets). The measurement indicates a redistribution of the momentum of jet constituents from smaller to larger radial distances from the jet axis in the presence of the QGP medium. These measurements provide new constraints for theoretical calculations of parton-flavor dependence of energy loss and jet--medium interactions in the QGP.

To understand the long-range ridge structures and the positive $v_2$ values observed for light-flavor particles in high-activity collisions of smaller systems as pp, p--Pb, and d--Au, the experiments also searched for non-zero $v_2$ signals in the heavy-flavor sector with the aim to provide insights whether the observed collectivity originates from initial- or final-state effects, or both. Experiments at the LHC performed several measurements to evaluate the $v_2$ of charm and beauty hadrons in pp and p--Pb collisions, using two-particle correlation techniques. A positive $v_2$ value for charm hadrons was measured, showing a rising and then decreasing trend with $p_{\rm{T}}$, and a mass ordering in the low-$p_{\rm{T}}$ region where heavier particles have smaller $v_2$ at a given $p_{\rm{T}}$. These trends are similar to those observed in Pb--Pb collisions, where they originate from parton interactions with the medium constituents described by hydrodynamic laws. The measurements in p--Pb collisions were compared with AMPT model calculations that generate positive $v_2$ via an anisotropic parton escape mechanism, and with CGC calculations that predict correlations in the initial stage of the collision between partons originating from projectile protons and gluons in the dense lead nucleus, resulting in a sizable $v_2$. Both models qualitatively provide compatible results and are able to describe the data, indicating that both initial- and final-state effects could explain the azimuthal anisotropies observed in small systems, while leaving the question of the exact origin of these effects still open. 

The measurements presented in this review were performed by the STAR and PHENIX Collaborations at RHIC, and by the ALICE, ATLAS, CMS, and LHCb Collaborations at the LHC, using data collected in 2018. While the current measurements have provided important information about various aspects of heavy-flavor production in hadronic collisions as discussed above, the uncertainties in several of these measurements are high, and extension to different kinematic regions are required to achieve the goals intended with these measurements. With the ongoing Run 3 data collection at the LHC, and the new sPHENIX experiment at RHIC, we can expect additional correlation measurements with improved precision, as well as access to further, more differential observables. These ongoing and future efforts will help to improve our knowledge of the abovementioned topics, and to shed further light on some of the open questions in the field. 

\vspace{6pt}
\authorcontributions{Conceptualization, F.C. and D.T.; investigation, F.C. and D.T.; writing---original draft preparation, F.C. and D.T.; writing---review and editing, F.C. and D.T. All authors have read and agreed to the published version of the manuscript.} 

\funding{This work was supported by U.S.
Department of Energy Office of Science under contract
number DE-SC0013391. } 

\conflictsofinterest{The authors declare no conflicts of interest.} 

\begin{adjustwidth}{-\extralength}{0cm}
\reftitle{References}

\PublishersNote{} 
\end{adjustwidth}

\end{document}